\documentclass[a4paper,12pt]{article}
\pdfoutput=1
\usepackage{graphicx}
\usepackage{epstopdf}
\usepackage{amssymb,amsmath}
\usepackage[colon,authoryear]{natbib}
\begin{document}

\title{Vertical instability and inclination excitation during planetary migration}

\author{G. Voyatzis, K. I. Antoniadou, K. Tsiganis\\
Department of Physics, Aristotle University of Thessaloniki, \\54124, Thessaloniki, Greece 
\\voyatzis@auth.gr, kyant@auth.gr, tsiganis@auth.gr }    
\maketitle
\begin{center}
 The final publication is available at Springer via\\ http://dx.doi.org/10.1007/s10569-014-9566-3 
\end{center}
\begin{abstract}
We consider a two-planet system, which migrates under the influence of dissipative forces that mimic the effects of gas-driven (Type II) migration. It has been shown that, in the planar case, migration leads to resonant capture after an evolution that forces the system to follow families of periodic orbits. Starting with planets that differ slightly from a coplanar configuration, capture can, also, occur and, additionally, excitation of planetary inclinations has been observed in some cases. We show that excitation of inclinations occurs, when the planar families of periodic orbits, which are followed during the initial stages of planetary migration, become vertically unstable. At these points, {\em vertical critical orbits} may give rise to generating stable families of $3D$ periodic orbits, which drive the evolution of the migrating planets to non-coplanar motion. We have computed and present here the vertical critical orbits of the $2/1$ and $3/1$ resonances, for various values of the planetary mass ratio. Moreover, we determine the limiting values of eccentricity for which the ``inclination resonance'' occurs.  
\end{abstract}
{\bf keywords} inclination excitation, type II migration, periodic orbits, vertical stability, planetary systems.

\section{Introduction}
In the last fifteen years, many studies have been devoted to planetary radial migration, induced by the interaction with the gaseous protoplanetary disk (commonly referred to as ``Type II'' migration, when the planetary masses are of the order of the Jovian mass). A common conclusion is that such migration leads with high probability to resonant capture and this may explain why many planets in extrasolar systems are found to be locked in resonance, see e.g.\ (\citealt{haghi99,leepeal02,np02,kley03,pap03}).

Planet migration is a complex process that cannot be viewed independently from the planet formation process. Some models suggest that capture in resonance can occur during the phase of "Type I" (fast)  migration, while the planets are still small in mass (embryos), (see e.g. \citealt{otto13}). However, planets that are nearly fully formed and have large masses can also be captured in resonance, during the subsequent phase of slow (Type II) migration, as has been seen in many simulations. For example, such a process (capture of Jupiter and Saturn in a 3/2 or 2/1 resonance) has been shown in simulations to produce the necessary initial set-up for the "Nice model" of solar system evolution (see \citealt{morc07}).

The dynamics of a two-planet system under a 'slow' dissipation have been studied in various approximations. In the secular approximation,  \citet{mr11} have shown that the two-planet system evolves along the corresponding stationary solutions of the secular equations, owing to the exchange of angular momentum between the planets and the external medium (disc); the {\it angular momentum deficit} (AMD) is an adiabatic invariant of the system. Similarly, near a resonance, a new action, $J$, conjugate to the resonant angle $\sigma$ can be defined  as an adiabatic invariant. The use of adiabatic invariance in problems of tidal migration and resonance crossing in the restricted three-body problem was pioneered by (\citealt{henr82,henrl83}) and generalized to the case of two massive satellites by \citealt{peale86}. As also shown in (\citealt{morbtgl09}) for the case of a two-planet system, when the critical curve of the resonance is crossed, $J$ suffers a very small jump and approximately is preserved, inside the resonance domain; the former action (e.g. the AMD) is no longer preserved. The system evolves along the resonant stationary solutions, moving from one energy level to the other, while the amplitude of oscillations around the equilibrium is nearly preserved. Note that each resonance is defined by a different critical argument and therefore, a different action that will serve as the new invariant, if several resonances are crossed under the action of a dissipative force. At each resonance crossing, the osculating elements suffer from instantaneous 'jumps' that are roughly independent of the crossing speed and can be large or small, depending on the geometry of the critical curve in phase space.

During resonant capture and as converging migration still proceeds, the eccentricities of the planets increase as a consequence of the preservation of an adiabatic invariant mentioned above. Presenting the evolution of the system in the eccentricities plane, we can obtain particular migration paths, that depend on the planetary mass ratio. These paths follow closely the stationary stable solutions of the averaged planetary three-body problem (as shown in \citealt{mebeaumich03,lee04,bmfm06}). Seen from a different point of view, these stationary solutions correspond to resonant, stable, periodic orbits of the general three body problem in a rotating frame. Therefore, the long-term evolution paths followed by a migrating two-planet system are described by families of stable periodic orbits (\citealt{hadjvoy10,hv11}). 

All the above mentioned studies assume coplanar planetary motion. Introducing a non-zero mutual inclination in a system, one could claim that the former stability of the system is not affected, since the average distance of the planets increases. However, this is not generally true and a small inclination may destabilize a planetary system, as in (\citealt{fmb05}). Regular evolution of inclined systems is expected in phase space islands around $3D$ stable families of periodic orbits (\citealt{av12}). \citet{thommes03} showed that $3D$ stable configurations can be obtained after migration and resonant capture in the $2/1$ resonance, starting from a two-planet system of nearly (but not exactly) coplanar orbits. In that work, it was observed that, when the system reached particular high-enough values of the eccentricity, an excitation of the mutual inclination took place; thus, the system reached a so-called  ``inclination resonance''. As long as the system remains nearly planar, it follows closely a path that is determined by the planar stationary solutions (which can be asymmetric). Numerical simulations (\citealt{leetho09}) showed that sudden ``jumps'' to nearby paths, with the outer planet typically being more eccentric than the inner one, can occur, leading to inclination excitation under particular circumstances. These paths can be identified as families of asymmetric periodic orbits (see e.g. \citealt{vkh09}). Considering reasonable values of the migration and eccentricity damping rates \citet{litsi09b} showed that capture to other resonances (e.g. $3/1$, $4/1$ and $5/1$) can also lead to inclination excitation, when eccentricity damping is not very strong. 

In this study, we show an intrinsic property of the three body problem dynamics that causes the inclination excitation and determine the particular regions in phase space, where it takes place. In particular, we show that {\em vertical instability} of periodic orbits along which a planetary system migrates, can excite the system away from its planar motion. Such a vertical instability occurs at vertical critical orbits (or briefly, {\em vco}) where families of $3D$ periodic orbits bifurcate. In the next section, we present a brief description of our model and its periodic orbits and discuss the notion of vertical instability. In Sect. \ref{SecVCO}, we compute and present the vertical stability property for the family of circular periodic orbits and for the $2/1$ and $3/1$ resonances. In Sect. \ref{SecMig}, we perform numerical simulations of a migrating two-planet system, which evolves along the circular family and then, is captured to either the $2/1$ or the $3/1$ resonance. We show that, for reasonable values of the migration rate, such two-planet systems can naturally reach non-zero values of mutual inclination, provided that the eccentricity damping rate is not very strong. Finally, we conclude and discuss the orbital features that such an inclined system should possess, if produced by differential migration.

\section{Periodic orbits and vertical instability}
We consider a two-planet system and study its dynamics in the general three body problem, consisting of a star, $S$, with mass $m_0$, and two planets, $P_1$ and $P_2$ with masses $m_i\ll m_0$, $i=1,2$. Indices 0, 1 and 2 will always indicate quantities of the star, the inner and the outer planet, respectively. By introducing a rotating frame of reference $Gxyz$, which rotates around the constant angular momentum vector and contains always the bodies $S$ and $P_1$ in the plane $Gxz$, the position of the system is determined  by the four variables $(x_1, x_2, y_2, z_2)$ (see \citealt{mich79,av12}). Thus, we obtain a four degrees of freedom Lagrangian   
\begin{equation} \label{Lagrangian}
\mathfrak{L}=\mathfrak{L}(\mathbf{q}, \mathbf{\dot q}),\quad \mathbf{q}=\{x_1,x_2,y_2,z_2\},
\end{equation}
and particularly,
$$
\mathfrak{L}=\frac{1}{2} (m_0+m_1)[a(\dot x_1^2+\dot z_1^2+x_1^2\dot \theta^2)+ b [(\dot x_2^2+\dot y_2^2+\dot z_2^2) 
+\dot\theta^2(x_2^2+y_2^2)+2\dot\theta(x_2\dot y_2-\dot x_2y_2)]]-V,
$$
where $m=m_0+m_1+m_2$, $a=m_1/m_0$, $b=m_2/m$, $V=-\frac{m_0 m_1}{r_{01}}-\frac{m_0 m_2}{r_{02}}-\frac{m_1 m_2}{r_{12}}$ is the potential with $r_{ij}$ indicating the distance between the bodies $i$ and $j$. The angle $\theta$, between the rotating and the inertial frame, is a cyclic variable, while variables $z_1$, $\dot z_1$ and $\dot \theta$ depend on the variables $\mathbf q$ (\citealt{av12}). 

For the system (\ref{Lagrangian}) we can define periodic solutions of period $T$, $\mathbf{Q}(T)=\mathbf{Q}(0)$, where $\mathbf Q=\{\mathbf{q}, \mathbf{\dot q}\}$. By studying the evolution on a Poincar\'e map, defined e.g. by the surface of section $y_2=0$ with $\dot y_2>0$, the set of initial conditions of periodic orbits 
$$
\mathbf{Q}(0)=\{x_{10},x_{20},y_{20}=0, z_{20},\dot x_{10},\dot x_{20}, \dot y_{20},\dot z_{20}\} 
$$
forms characteristic curves (or families of periodic orbits) in the phase space of the Poincar\'e map, due to mono-parametric continuation.
Linear stability analysis of a periodic orbit is based on the position of the four pairs of  conjugate eigenvalues of the monodromy matrix (\citealt{skokos01}). 

Planar periodic orbits can be computed in the context of the planar model ($z_2=\dot z_2=0$) and their linear stability is characterized as {\em horizontal} stability. 

The property of {\em vertical} stability of planar periodic orbits has been introduced by \citet{hen} for the restricted problem and generalized for the general problem by \citet{mich79} and \citet{ikm78}.  

Let us consider a planar solution $\bar{\mathbf{Q}}(t)$, which corresponds to the initial conditions $\bar{\mathbf{Q}}(0)$ with $z_2(0)=\dot z_2(0)=0$. We expand the equation of motion for $z_2$, as it is derived by the Lagrangian (\ref{Lagrangian}), around the planar solution assuming a small vertical deviation in the initial conditions $\delta z_2(0)=\zeta(0)$ and $\delta\dot z_2(0)=\eta(0)$. Then, to first order in terms of the vertical components, the evolution of the deviations is given by the equations  
\begin{equation} \label{Vdeq}
\dot\zeta=\eta,\quad \dot \eta=A(\bar{\mathbf{Q}}(t))\zeta+B(\bar{\mathbf{Q}}(t))\eta.
\end{equation}   
 
If $\bar{\mathbf{Q}}(t)$ is a periodic solution of period $T$, then (\ref{Vdeq}) is a linear system with periodic coefficients and solution of the form
\begin{equation}\label{LinSol}
\left ( \begin{array}{c} \zeta(t) \\ \eta(t)\end{array} \right )= \left ( \begin{array}{cc} \zeta_1(t) & \zeta_2(t) \\ \eta_1(t) & \eta_2(t) \end{array} \right ) \left( \begin{array}{c} \zeta(0) \\ \eta(0)\end{array} \right ) 
\end{equation}
where $(\zeta_1,\eta_1)^\top$ and $(\zeta_2,\eta_2)^\top$ forms the fundamental matrix of solutions $\mathbf{\Delta}(t)$ corresponding to the initial deviations $(1,0)^\top$ and $(0,1)^\top$ (\citealt{ikm78}). The evolution of solutions (\ref{LinSol}) depends on the eigenvalues of the monodromy matrix 
$$
\mathbf{\Delta}(T)=\left ( \begin{array}{cc} \zeta_1(T) & \zeta_2(T) \\ \eta_1(T) & \eta_2(T) \end{array} \right )=\left ( \begin{array}{cc} a & b \\ c & d \end{array} \right ),
$$
or on the {\em vertical stability index}
\begin{equation}
a_v=\frac{1}{2}(a+d).
\end{equation}
In particular, if $|a_v|<1$, the eigenvalues are complex conjugate with modulus 1 and the vertical motion is stable (bounded). If $|a_v|>1$ the eigenvalues are real, leading to unbounded solutions (vertically unstable). Periodic orbits with $|a_v|=1$ are {\em vertical critical orbits} ({\em vco}) and, in general, constitute critical planar orbits from which families of $3D$ periodic orbits bifurcate (\citealt{ichmich80,av12}).    

\begin{figure}[t]
\begin{center}
\includegraphics[width=12cm]{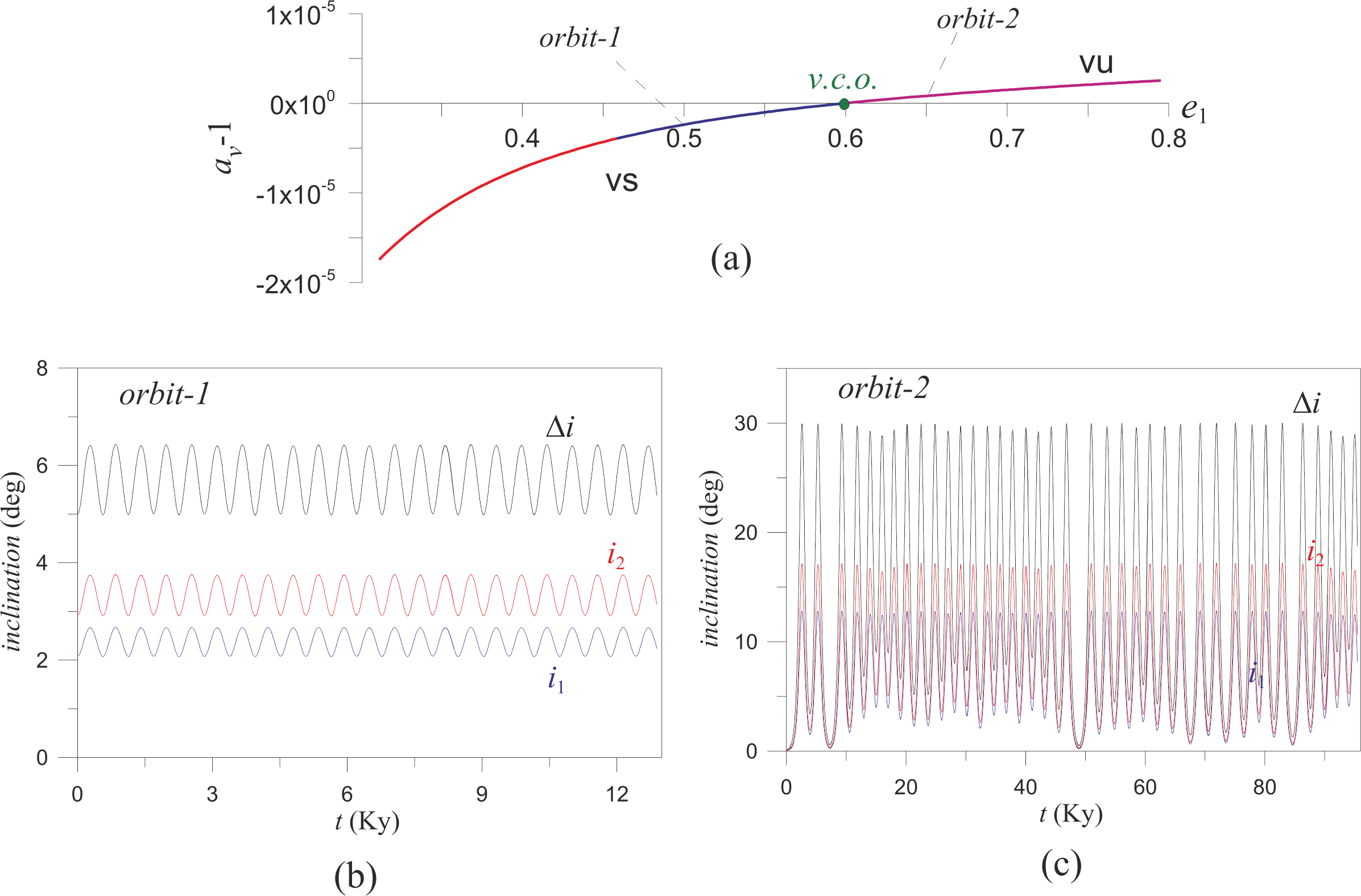} 
\end{center}
\caption{{\bf a} The variation of stability index along the $2/1$ resonant family of planar periodic orbits for $\rho=0.5$ (family is parametrized by $e_1$). The {\em vco} is located at $e_1=0.6$ ($e_2=0.27$). The indicated {\em orbit-1} ($e_1=0.5$, $e_2=0.21$) and all orbits on the left of {\em vco} are vertically stable (vs); {\em orbit-2} ($e_1=0.65$, $e_2=0.3$) and all orbits on the right of {\em vco} are vertically unstable (vu). The red coloured section corresponds to horizontally unstable orbits {\bf b} The evolution of planetary inclinations of the planar {\em orbit-1} that is initially perturbed vertically by $\Delta i(0)=5^\circ$. {\bf c} The same for {\em orbit-2} initially perturbed by $\Delta i(0)=0.1^\circ$.}
\label{FigVsuEvol}
\end{figure}

\begin{figure}
$
\begin{array}{cc}
\includegraphics[width=6cm]{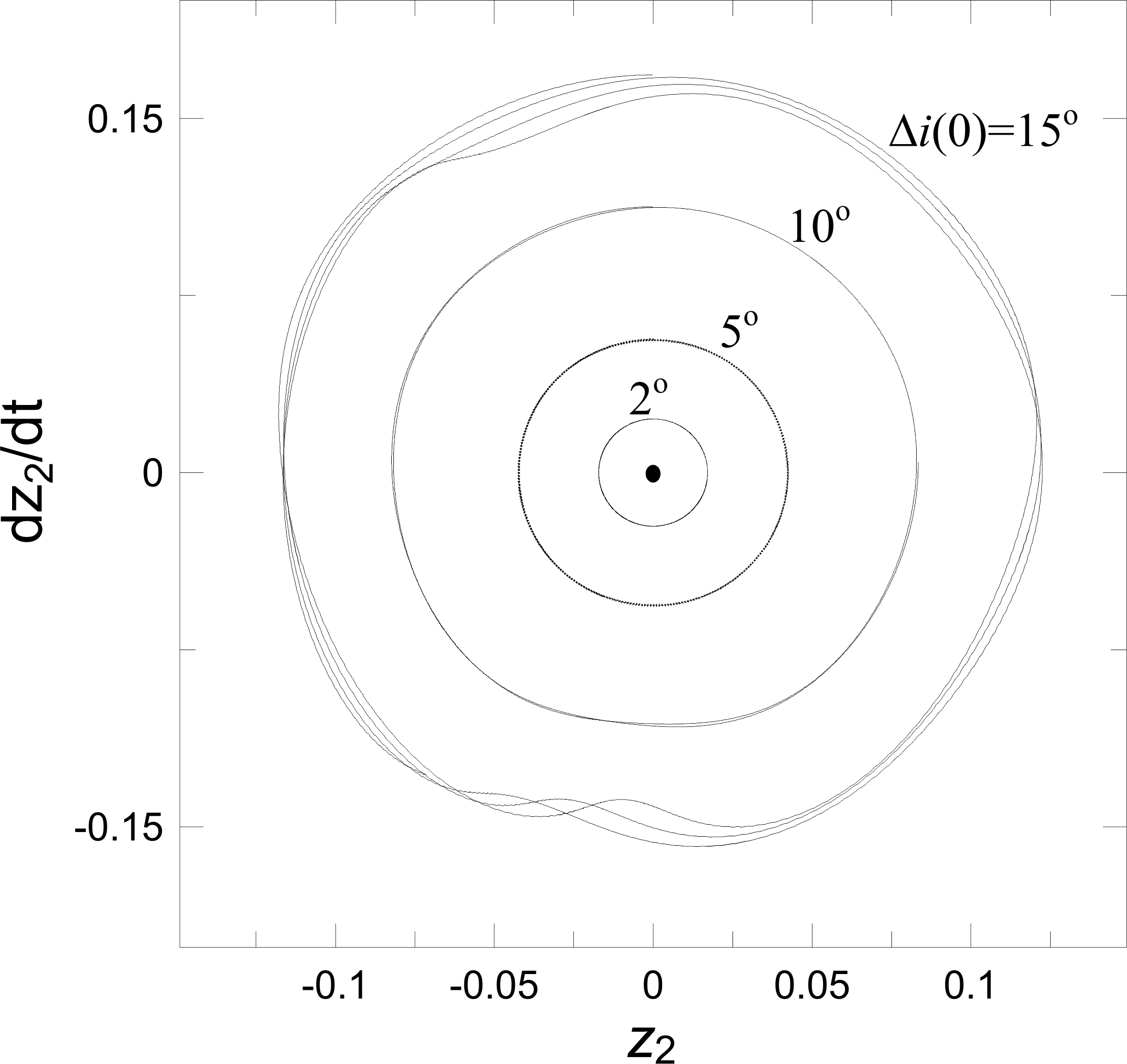}  & \includegraphics[width=6cm]{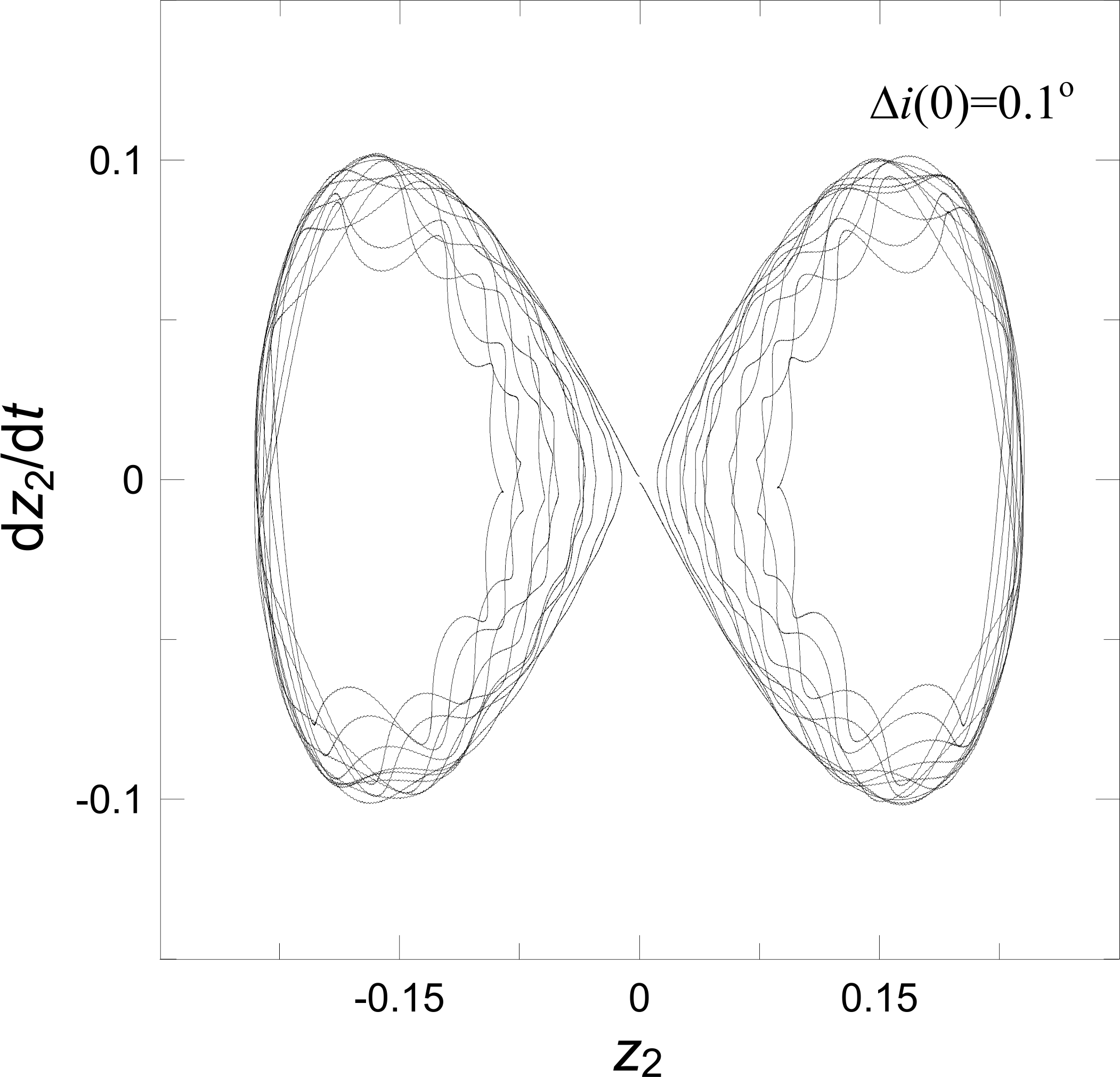} \\
\textnormal{(a)} &\textnormal{(b)}  
\end{array}$
\caption{Poincar\'e map projections in the plane $(z_2,\dot z_2)$ of {\bf a} vertically stable orbits ({\em orbit-1}) starting with a vertical deviation from plane given by the values $\Delta i(0)$ {\bf b} two vertically unstable orbits starting with the same initial planar conditions ({\em orbit-2}) and with a very small initial vertical deviation $\Delta i(0)$=$0.1^\circ$. The two orbits have been chosen so that they follow the unstable manifolds of both directions.} 
\label{FigPncV}
\end{figure}

In general, $a_v$ varies along a family of planar periodic orbits and we may obtain none, one or more {\em vco} along the family. In Fig. \ref{FigVsuEvol}a, we present the variation of $a_v$ along a part of the $2/1$ resonant symmetric family (see also Fig. \ref{FigEfam21}a) for mass ratio $\rho=m_2/m_1=0.5$ and configuration $(\theta_1,\Delta \varpi)=(0,0)$, where $\theta_1=\lambda_1-2\lambda_2+\varpi_1$ is a resonant angle and $\Delta \varpi=\varpi_2-\varpi_1$ is the relative longitude of the periapse. Here, the family is parametrized by the eccentricity $e_1$. We obtain one {\em vco} at $(e^*_1,e^*_2)$=$(0.6,0.27)$. Orbits with $e_1<e^*_1$ or $e_1>e^*_1$ are vertically stable or unstable, respectively. By choosing initial conditions of a vertically stable orbit, e.g.\ a planar periodic orbit of the family at $(e_1,e_2)$=$(0.5,0.21)$, and by adding a small initial vertical deviation corresponding to a mutual planetary inclination $\Delta i(0)=5^\circ$ (and $\Delta \Omega = 180^\circ$) we obtain  regular, small-amplitude oscillations of the planetary inclinations (Fig. \ref{FigVsuEvol}b). Instead, if we consider the initial conditions of the planar periodic orbit $(e_1,e_2)$=$(0.65,0.3)$ and add a very small vertical deviation ($\Delta i(0)=0.1^\circ$), we obtain an excitation of the planetary inclinations, which now oscillate irregularly with relatively large amplitude. Also, we depict the above samples of trajectory evolution in the projection plane $(z_2,\dot z_2)$ of the Poincar\'e map (Fig. \ref{FigPncV}). In the case of vertical stability (left panel), we see that, even for large initial vertical deviations (up to $15^\circ$ of mutual inclination), the evolution is regular, showing vertical oscillations around the planar periodic orbit. In the case of vertical instability, a saddle-like picture is obtained, showing relatively large vertical deviations from the plane $z=0$.

\section{Vertical critical orbits in planetary dynamics} \label{SecVCO}
In this section, we compute and present {\em vco} in three cases, particularly for circular planetary orbits, $2/1$ and $3/1$ resonant orbits. The planar families of these orbits and their horizontal stability have been given in previous works, as it is mentioned for each case in the following. 

\begin{figure}
$$
\begin{array}{cc}
\includegraphics[width=6cm]{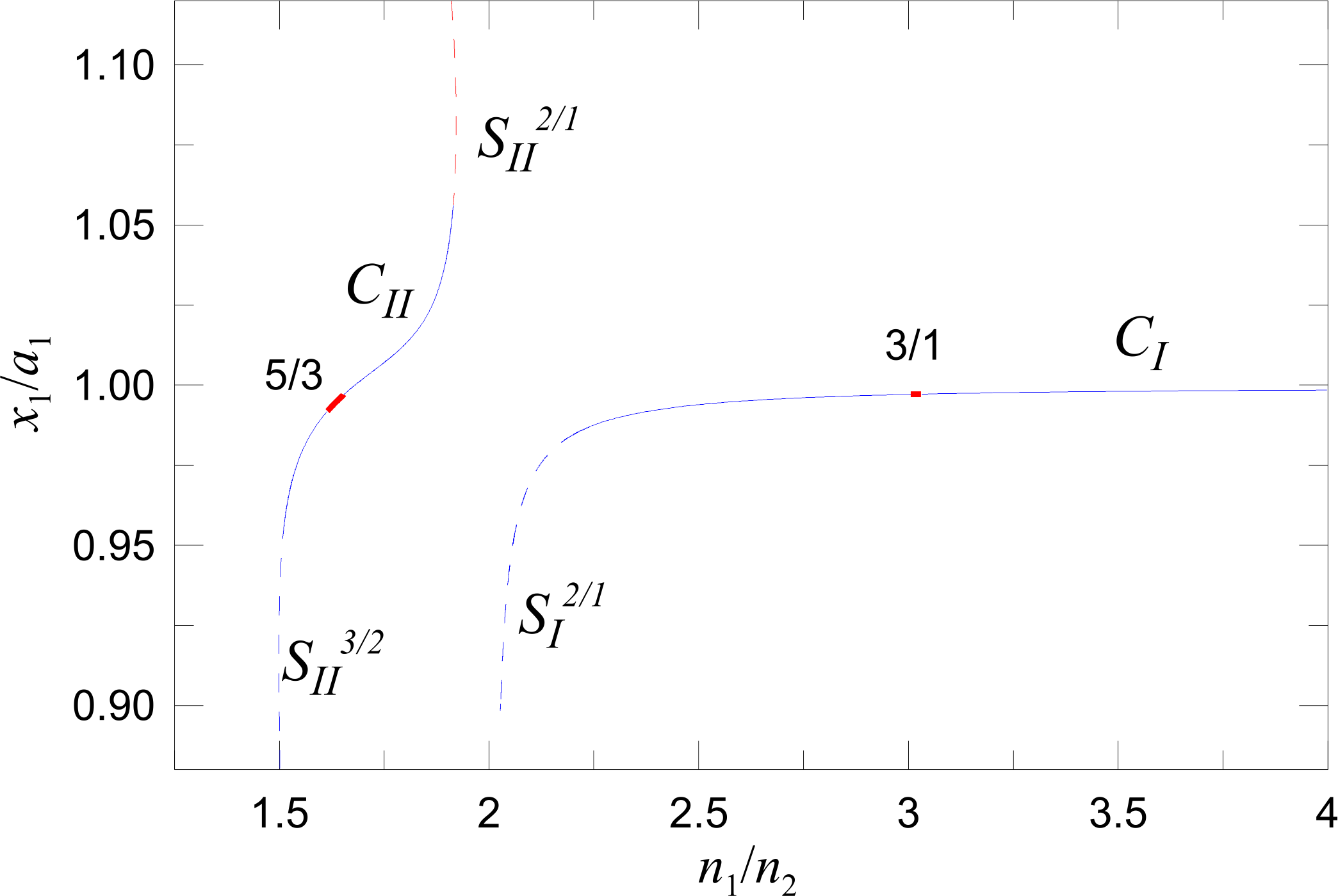}  & \includegraphics[width=6cm]{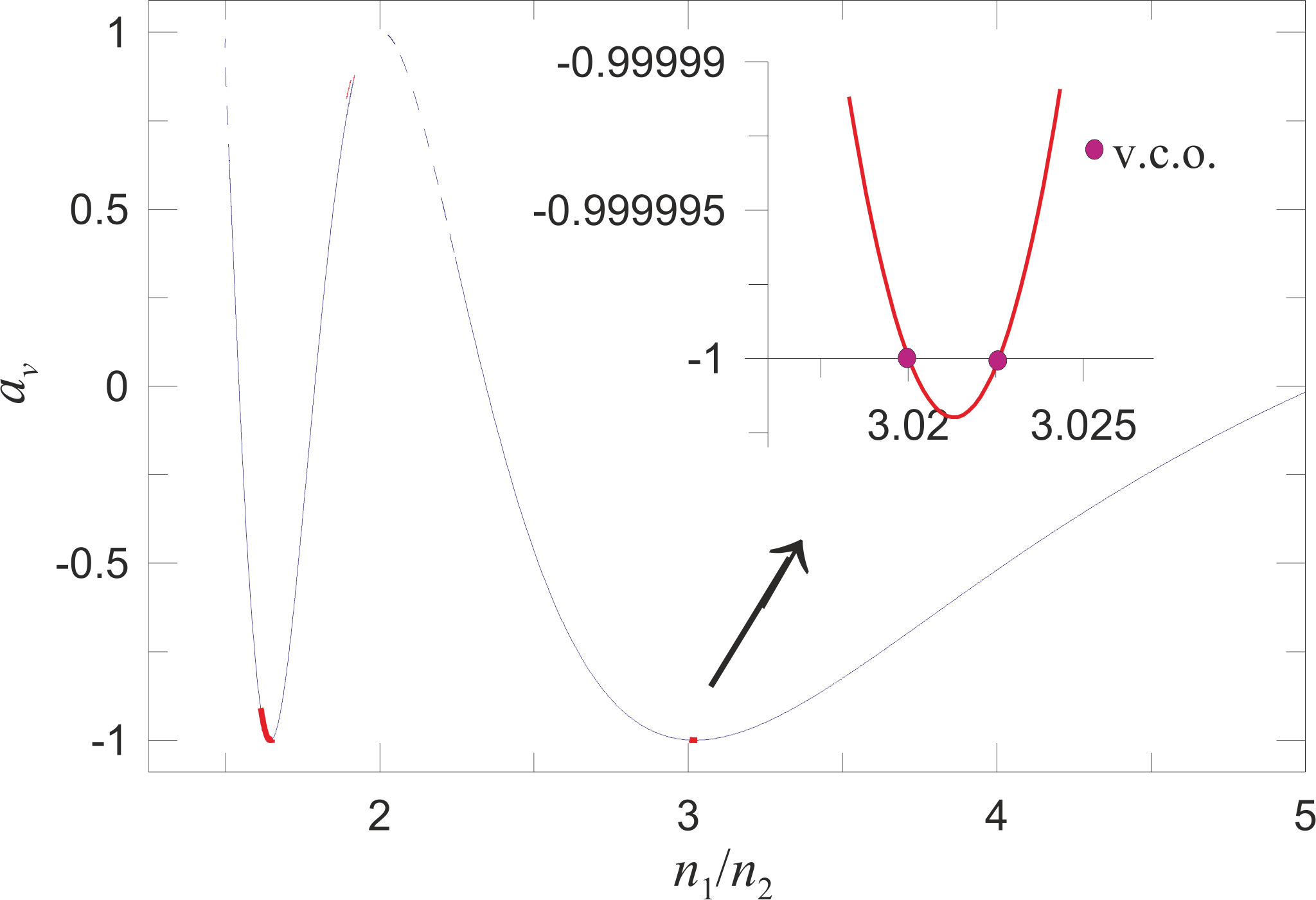} \\
\textnormal{(a)} &\textnormal{(b)}  
\end{array}$$
\caption{{\bf a} A presentation of the circular family segments $C_I$ and $C_{II}$ (solid lines) for $m_1=0.001$ and $m_2=0.002$. Horizontal axis indicates the mean motion ratio and the vertical axis the $x_1$ variable of the rotating frame, normalized s.t. $a_1=1$. Blue and red colour indicate stable and unstable orbits, respectively. Dashed curves indicate the resonant elliptic families {\bf b} The vertical stability index $a_v$ along the circular family. A magnification near the $3/1$ resonance is included.} 
\label{FigCircFam}
\end{figure} 

\subsection{The circular family}\label{SecCircFam}
The circular family of periodic orbits of the restricted problem can be continued with respect to the mass in the general problem (\citealt{hadj06,vkh09}) without structural changes, for small planetary masses. The mean motion ratio $n_1/n_2$ varies along the circular family, which shows gaps at the 1st order resonances $(n+1)/n$, $n\in {\mathbb Z}$. For $n_1/n_2\gtrsim 2/1$ we have the family segment $C_I$ (see Fig. \ref{FigCircFam}a), which consists of stable orbits except for a small section at the $3/1$ resonance, where the circular orbits become unstable. At the $2/1$ resonance the family continues smoothly to the elliptic $2/1$ resonant family $S_I^{2/1}$. For $n_1/n_2\lesssim 2/1$ the family segment $C_{II}$ exists, which continues smoothly to the elliptic resonant families $S_{II}^{2/1}$ and $S_{II}^{3/2}$ (to the right and to the left, respectively). A small horizontally unstable section of $C_{II}$ appears close to the $5/3$ resonance. 

The variation of the vertical stability index, $a_v$, along the circular family is shown in Fig. \ref{FigCircFam}b. At the segment $C_I$ we obtain that $a_v<-1$ in a small segment near the $3/1$ resonance. Thus, both the {\em vco}'s ($a_v=-1$) and the vertically unstable orbits are horizontally unstable. A similar situation holds for $C_{II}$ at the $5/3$ resonance. 
       
Similar results with those presented in Fig. \ref{FigCircFam} have been obtained for various mass ratios and for values of the planetary masses up to $5~M_J$, where  $M_J$ is the Jovian mass. 

\subsection{The $2/1$ resonance}
\begin{figure}
$$
\begin{array}{cc}
\includegraphics[width=6cm]{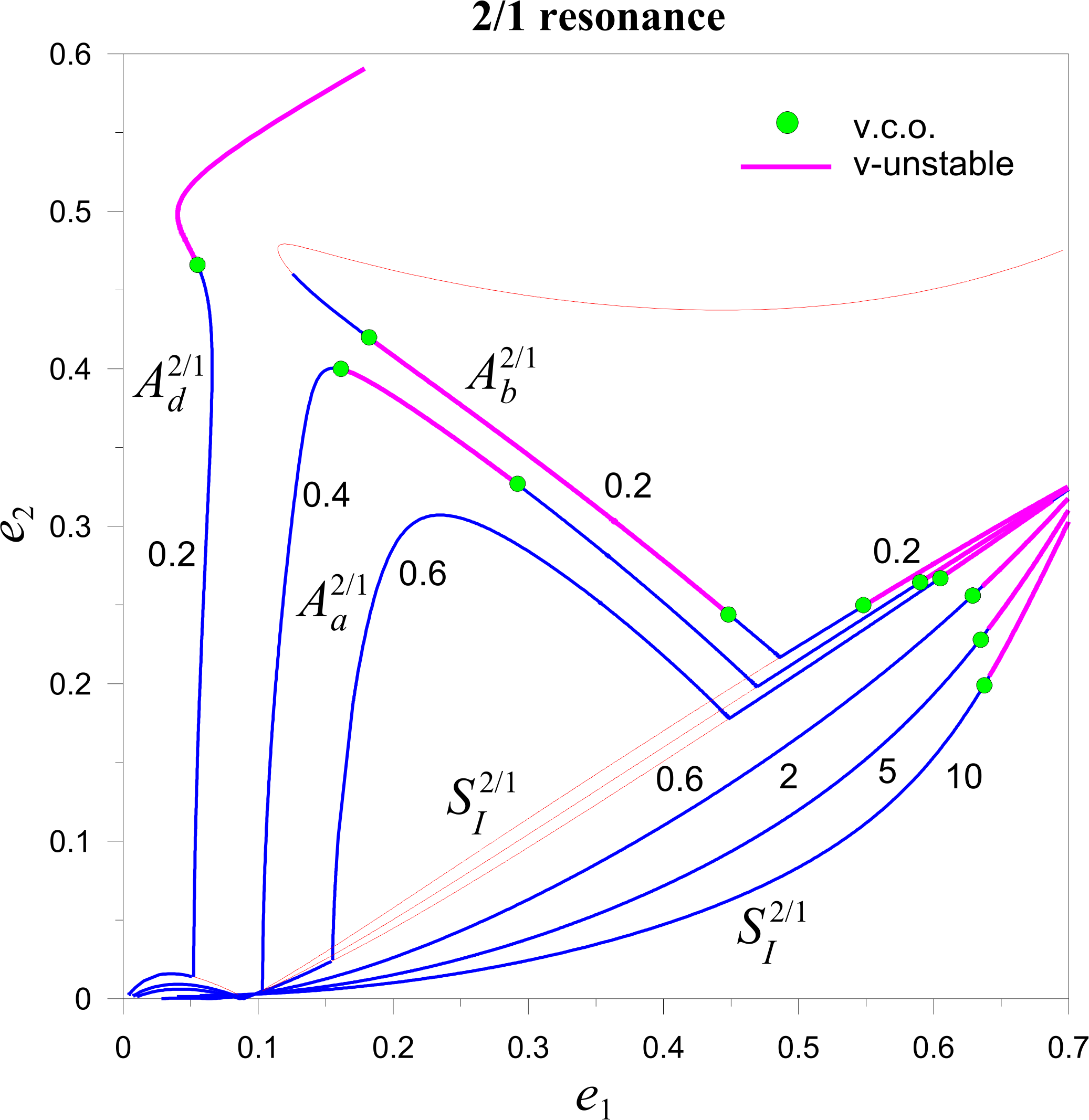}  & \includegraphics[width=6cm]{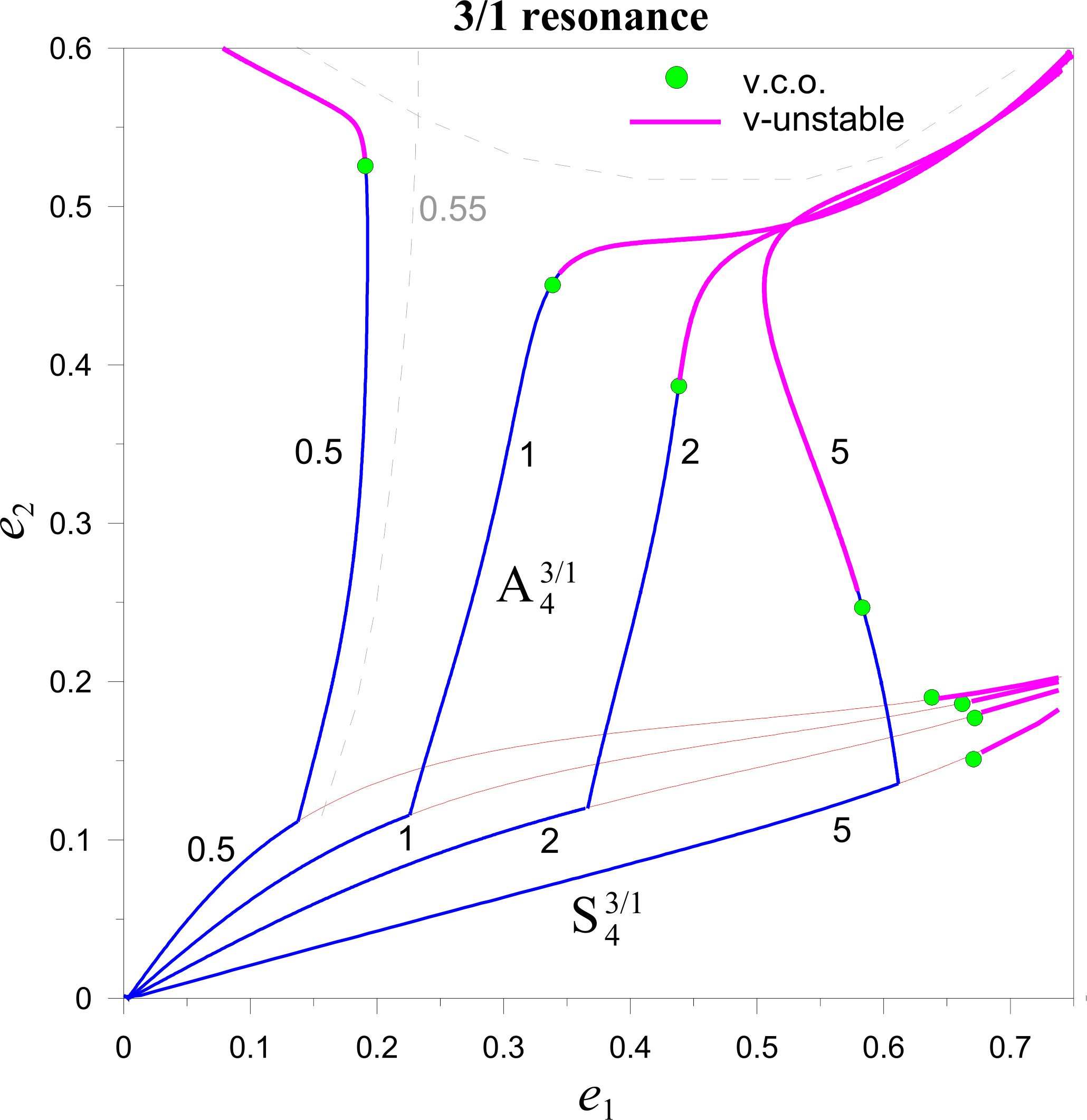} \\
\textnormal{(a)} &\textnormal{(b)}  
\end{array}$$
\caption{Families of {\bf a} $2/1$ and {\bf b} $3/1$ resonant periodic orbits. Blue and red colour indicate horizontal stability and instability, respectively. The circles indicate {\em vco} and the sections of magenta colour consist of vertically unstable orbits. For each curve the corresponding mass ratio $\rho=m_2/m_1$ is indicated.}
\label{FigEfam21}
\end{figure}

As mentioned above, from the circular family we obtain the resonant families $S_I^{2/1}$ and $S_{II}^{2/1}$ of symmetric periodic orbits. The family $S_{II}^{2/1}$ is horizontally unstable and terminates at a collision orbit. Thus, we examine only family $S_I^{2/1}$ for various values of the mass ratio $\rho=m_2/m_1$, as depicted in Fig. \ref{FigEfam21}a, in the projection plane of eccentricities. The properties of these families are described in \citet{bmfm06}, \citet{voyhadj05} and \citet{vkh09}. The families start from the circular family, $(e_1,e_2)\approx (0,0)$ as stable, but for $\rho\lesssim 1$ they turn into horizontally unstable at particular eccentricity values. Hence, we obtain an unstable segment which terminates at higher eccentricities and the orbits become stable again. At the points where the stability changes, families of asymmetric periodic orbits bifurcate. For $0.37\lesssim\rho\lesssim 1$, we obtain the family $A_a^{2/1}$, which forms a bridge connecting the two bifurcation points of the family $S_I^{2/1}$. For $\rho<0.37$, a bifurcation occurs and we obtain the asymmetric families $A_d^{2/1}$ and $A_b^{2/1}$.

We have computed and present here (also in Fig. \ref{FigEfam21}a) the vertical critical orbits and the family segments that are vertically unstable. In family $S_I^{2/1}$ there exist horizontally stable {\em vco}'s for $e_1>0.5$ and $\rho>0.12$. In family $A_a^{2/1}$, we obtain a bifurcation of {\em vco} at the critical mass ratio value $\rho^*\approx 0.43$. Thus, for $\rho>\rho^*$ no {\em vco} exists on the asymmetric family, which is always vertically stable. For $\rho<\rho^*$ two {\em vco}'s exist in $A_a^{2/1}$, which form a family section of vertically unstable orbits between them. These {\em vco}'s seem to continue also in family $A_b^{2/1}$. In family $A_d^{2/1}$, we have found one {\em vco} located at a low (resp.\ high) eccentricity value for the inner (resp.\ outer) planet.

\subsection{The $3/1$ resonance}
It has been shown that four $3/1$ resonant families of periodic orbits bifurcate from the circular family $S_I^{3/1}$ (\citealt{vh06,mbf06,voyatzis08}). From these families, only family $S_4$ starts from the circular family as stable for all values of the mass ratio $\rho$. Its characteristic curves in the eccentricities plane are shown in Fig. \ref{FigEfam21}b, for some typical values of $\rho$. For any value of $\rho$, the family becomes horizontally unstable at a particular eccentricity value, which depends on $\rho$. At these points bifurcation of the asymmetric family $A_4^{3/1}$ occurs, consisting of stable orbits (at least in the eccentricities domain of Fig. \ref{FigEfam21}b).    

The {\em vco}'s in the above mentioned families and the vertically unstable parts are also indicated in Fig. \ref{FigEfam21}b. We can see that symmetric {\em vco}'s are all horizontally unstable up to $\rho\approx 6.5$ \footnote{For $\rho\gtrsim6.5$, the {\em vco} enters the stable segment of $S_4^{3/1}$.} and located at $e_1\gtrsim 0.6$. Family $A_4^{3/1}$ has, also, one {\em vco} at a given point in $(e_1,e_2)$; as $\rho$ increases, $e_1$ also increases, while $e_2$ decreases.
 

\section{Vertical instability during planetary migration} \label{SecMig}

As we mentioned in the introduction, planetary migration, caused by planet-disc interactions, will force a nearly circular, planar two-planet system to evolve in phase space along the stable families of periodic orbits. \citet{leetho09} showed that this, also, holds for systems starting with slightly inclined orbits, but an inclination resonance may occur at some point. 

In this section, we wish to study the time evolution of a nearly (but not exactly) coplanar system of two planets, under the effects of radial (presumably, gas-driven) migration. To do this, we consider the planetary three-body problem, but impose a Stokes-type dissipative force on the outer planet that mimics the effects of Type II migration (\citealt{bf93,bmfm06}) of the form
\begin{equation} \label{Fdissipative}
\mathbf{F}_d=-C({\bf{\upsilon}}_p-\alpha \mathbf{\upsilon}_c)
\end{equation}
where $\mathbf{\upsilon}_p$ is the planar velocity component of the planet and $\mathbf{\upsilon}_c$ is the circular velocity at the particular distance of the star. The positive constants $C$ and $\alpha$ are associated in a first order approximation with the migration rate in semi-major axis, $\nu$, and the eccentricity damping, $K$, according to the formulae (\citealt{bf93,bmfm06})
$$
\nu=2C(1-\alpha),\quad K=\frac{\alpha}{2(1-\alpha)}.
$$
We performed a series of numerical simulations, with star mass $m_0=1 M_\odot$, starting with almost circular and co-planar orbits, i.e.\ $e_1(0)=e_2(0)=0.01$, $i_1(0)=i_2(0)=0.1^\circ$, and $\Delta\Omega=180^\circ$. The inner planet is always set to $a_1(0)=5~$AU. We examine two cases for the initial position of the outer planet (i) $2/1<n_1/n_2<3/1$ (interior to the $3/1$ resonance) and (ii) $n_1/n_2>3/1$ (exterior to the $3/1$ resonance). Also, following \citet{litsi09b}, where the formula concerning the order of migration rate given by \citet{war97} is used, we consider 
parameter values in the intervals $10^{-7} \leq \nu \leq 10^{-5}$ $(y^{-1})$ and $0.5\leq K\leq 100$. However, the typical dynamics presented in the following is revealed for small values of migration rates, particularly for about $\nu\lesssim 10^{^-6} y^{-1}$, and for sufficiently small eccentricity damping $K$, in order for the system to reach the necessary eccentricity values. 

Since we always start with orbits close to the circular family $C_I$, the dissipative forces cause a slow slide of the system along the family $C_I$ (i.e.\ the planetary orbits remain nearly circular) and towards lower mean motion ratio values (inward migration). When we start below the $3/1$ resonance, the system enters the family $S_I^{2/1}$ and is captured in the $2/1$ resonance. Starting above the $3/$1 resonance, we approach the $3/1$ resonance where the $C_I$ becomes horizontally unstable and contains also vertical critical points. However, the system slides into the $S_4^{3/1}$ resonant family, which is the only stable one, and $3/1$ resonant capture occurs. During the migration of the system along the family $C_I$, for both cases, we have vertical stability and therefore, the orbits remain almost planar. We note that, apart from $2/1$ and $3/1$, capture to other resonances has not been observed in our set of simulations. 

\begin{figure}
\centering
\includegraphics[width=12cm]{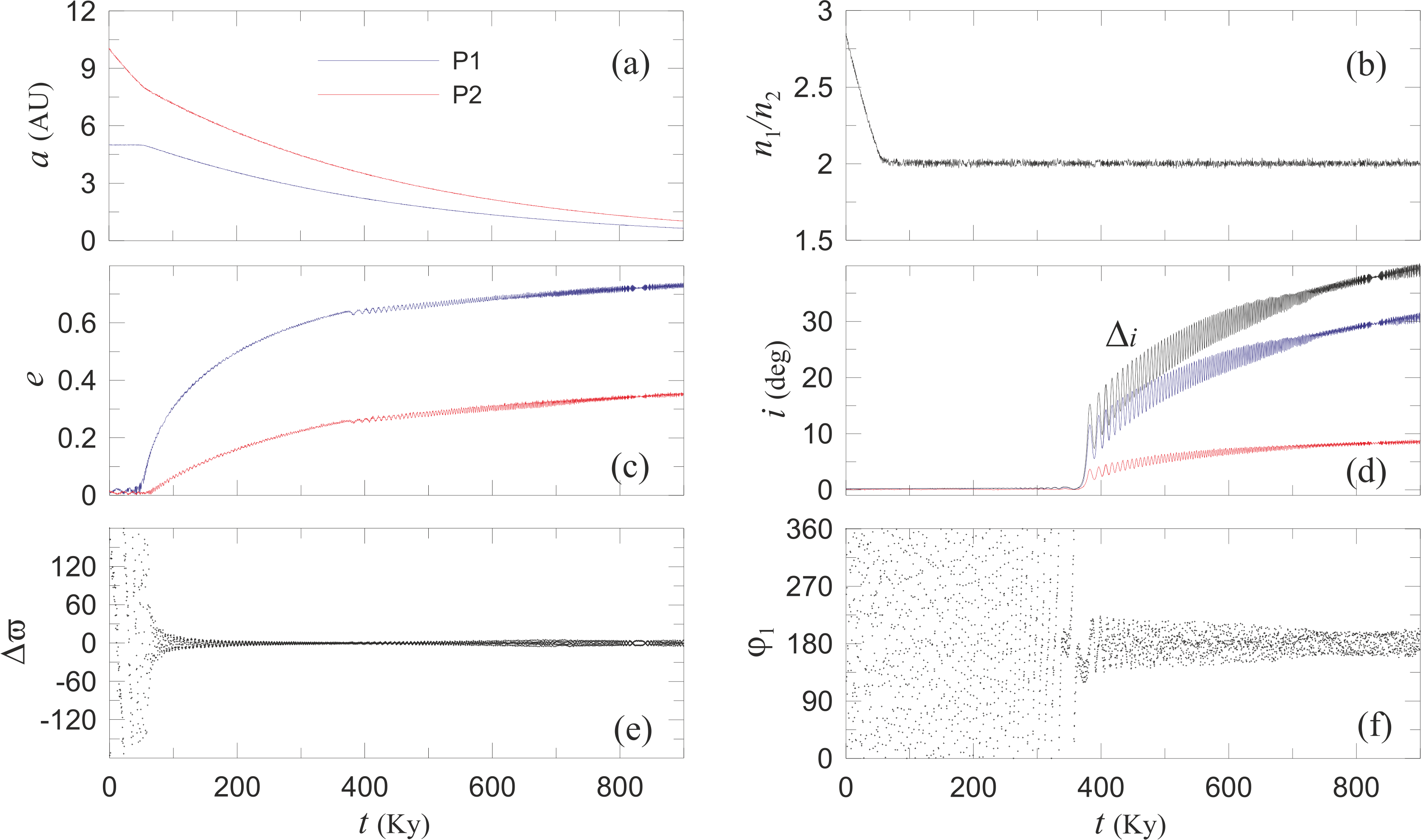}
\caption{Evolution of orbital elements under the influence of the dissipative force (\ref{Fdissipative}) with $\nu=4.2~10^{-6}\textnormal{y}^{-1}$, $K=1$ and planetary masses $m_1=1~M_J$, $m_2=2~M_J$. Blue and red colour lines refer to the inner ($P_1$) and the outer planet ($P_2$), respectively.  {\bf a} semimajor axes {\bf b} the mean motion ratio {\bf c} eccentricities {\bf d} inclinations and mutual inclination, $\Delta i$ {\bf e} the apsidal difference $\Delta\varpi$ {\bf f} the resonant angle $\phi_1$ (similar evolution is observed for $\phi_2$)} 
\label{FigMigExOE}
\end{figure}    

\begin{figure}
\centering
\includegraphics[width=8cm]{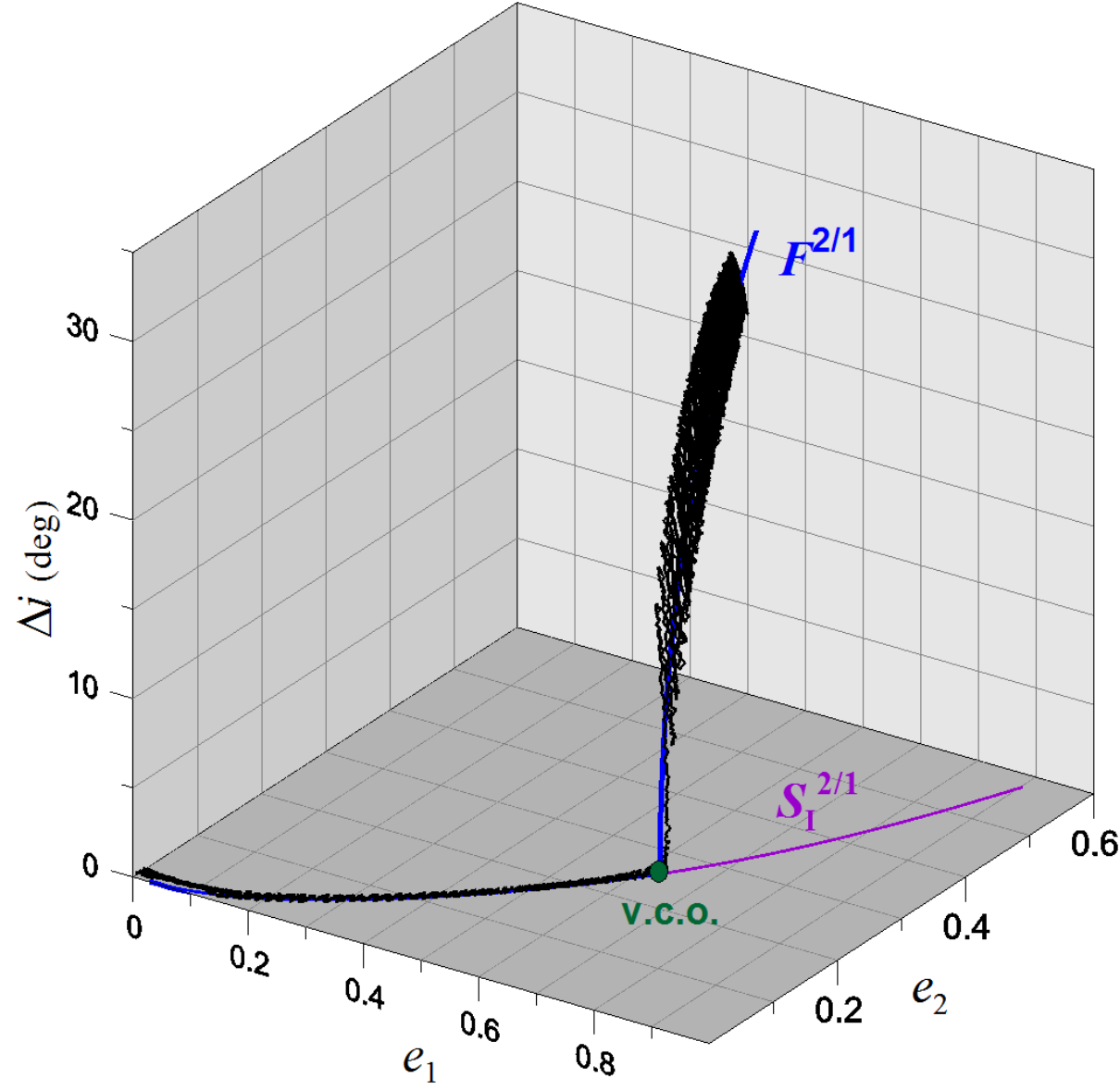}
\caption{The evolution shown in Fig. \ref{FigMigExOE} is now presented in the space $e_1-e_2-\Delta i$. When the system is captured in the resonance $2/1$ (approximately at zero), it evolves along the path indicated by the family $S_I^{2/1}$. The eccentricities increase, while the inclinations remain at low values. When the {\em vco} is reached the inclinations start to increase and the system follows the family $F^{2/1}$ of $3D$ periodic orbits.} 
\label{FigMigEx3D}
\end{figure}

\subsection{$2/1$ resonance capture}

After $2/1$ resonant capture occurs, the system follows the family $S_I^{2/1}$. If the system has mass ratio $\rho>1$, it does not leave the symmetric family and reaches the corresponding {\em vco}, located at $e_1>0.5$. A typical example of such an evolution is shown in Fig. \ref{FigMigExOE} for $\rho=2$, $\nu=4.2~10^{-6}\textnormal{y}^{-1}$ and $K=1$.  Resonant capture takes place at $t\approx 40$Ky, with the eccentricities remaining low up to that time. After getting captured, the eccentricities of the planets start increasing, but the orbits still remain planar. At $t\approx 360$Ky, the inclination resonance is reached and the planetary orbits become mutually inclined. At the inclination resonance, we have $e_1=0.63$ and $e_2=0.26$; these values correspond to the position of the {\em vco} of the family $S_I^{2/1}$. As the inclination starts to increase the particular resonant angles $\varphi_i$=$2\lambda_1-4\lambda_2+2\Omega_i$ ($i=1,2$) librate around $180^\circ$ indicating, beside the libration of $\Delta\varpi$ around $0^\circ$ the symmetric configuration of the system. 

It is clear from Fig. \ref{FigMigExOE} that the system evolves in a way similar to that discussed in the Introduction. When the critical mean motion ratio 2/1 is reached along the circular family (at $t\approx 40~$Ky), a gap is met and the system turns from the family of circular periodic orbits to the 2/1-resonant family of eccentric periodic orbits (see Sect. \ref{SecCircFam}). The AMD is no longer a proper invariant; the resonant action (conjugate to the resonant angle) is the invariant that will be approximately preserved inside this particular resonance domain. When the critical curve of the 2/1-inclination resonance is reached (i.e.\ the {\it vco}) a similar pattern is observed due to the bifurcation of the 3D family of periodic orbits. The new action (conjugate to the critical angle of the 2/1-inclination resonance) will be now preserved; for $t > 400~$Ky, the inclinations increase as migration goes on. The only appreciable difference between the two resonance crossings that the system suffers is in the initial 'jump' in osculating elements. We mention that in the planar problem the passage from the circular family to the elliptic resonant family is quite smooth (due to the formed gap, see Fig. \ref{FigCircFam}), but in the passage from the planar to the 3D family we have a bifurcation point at the {\it vco} that causes an abrupt change in the inclination. Also, one can see in Fig. \ref{FigPncV}b, that the area enclosed by the 'separatrix' is much wider than before this bifurcation occurs (where no separatrix exists), and thus the change in osculating inclination upon the {\it vco} crossing has to be large, if the system has initially nearly zero inclinations.

If we present the evolution in the space $e_1-e_2-\Delta i$ (see Fig. \ref{FigMigEx3D}), we observe that the inclination starts to increase, when the {\em vco} is reached and the family becomes, thereafter, vertically unstable. In particular, we observe that the evolution follows the family $F^{2/1}$ of $3D$ orbits, which bifurcates from the {\em vco}. As seen from the nearly vertical intersection between the planar and $3D$ families, crossing of the {\it vco} implies that small variations in eccentricity will be accompanied by large variations in inclination.

For $0.43<\rho<1$, the evolution follows the asymmetric branch $A_a^{2/1}$, when $S_I^{2/1}$ becomes horizontally unstable (see Fig. \ref{FigEfam21}a). Since $A_a^{2/1}$ ends again at the stable part of $S_I^{2/1}$, the evolving system meets again the {\em vco} of the symmetric family at $e_1>0.5$. The excitation of inclination occurs again, when the {\em vco} is reached.

We remind the reader that for mass ratios $\rho<\rho^*$, where $\rho^*=0.43$, vertical critical orbits exist along the family $A_a^{2/1}$. Thus, the inward migration for $0.37<\rho<\rho^*$ may lead to the {\em vco} of this family, which is located at a relatively low eccentricity value, $e_1$ ($\lesssim 0.2$). For $\rho<0.37$, the system follows the asymmetric family $A_d^{2/1}$, which has, also, a {\em vco} for $e_1<0.2$, but for relatively large eccentricity of the outer planet (namely, $e_2>0.4$). 
        
\begin{figure}
$$
\begin{array}{cc}
\includegraphics[width=6cm]{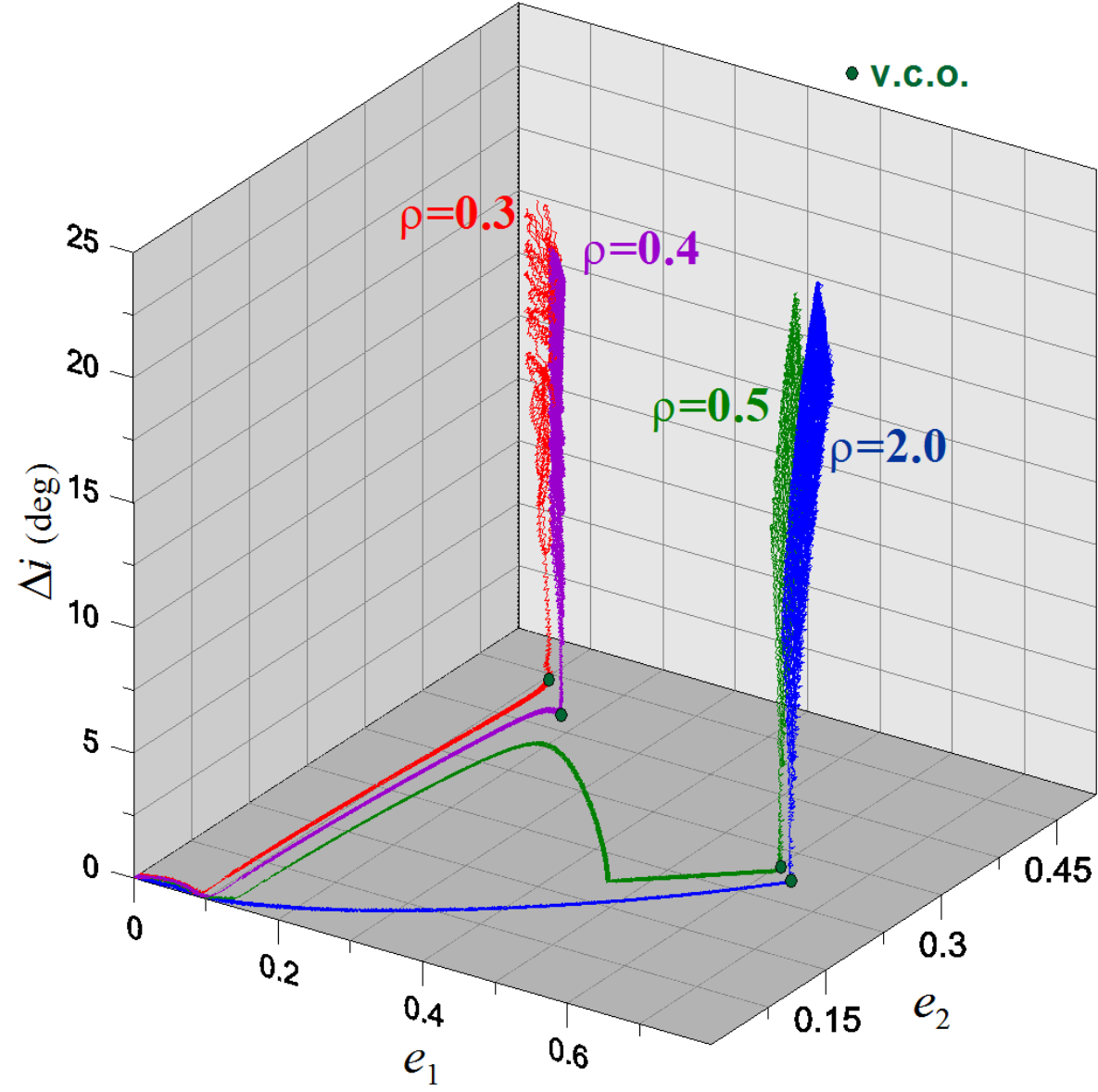}  & \includegraphics[width=6cm]{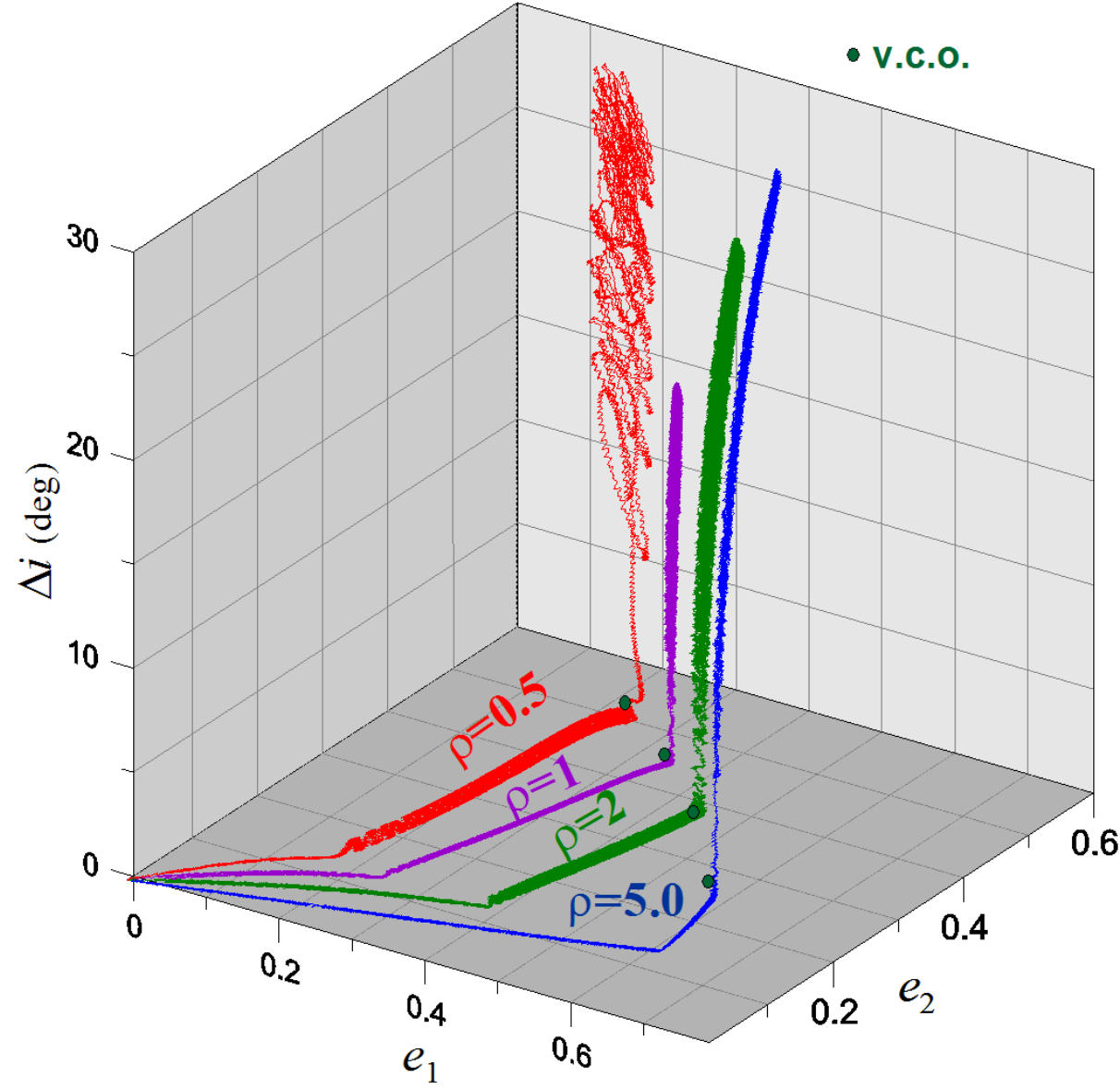} \\
\textnormal{(a)} &\textnormal{(b)}  
\end{array}$$
\caption{The evolution after \textbf{a} $2/1$ and \textbf{b} $3/1$ resonant capture presented in the space $e_1-e_2-\Delta i$. Four different cases are presented, which correspond to the indicated planetary mass ratio values. It is $\nu=4.2~10^{-7}~y^{-1}$ and $K=1$ (except for $\rho=0.3$ where $K=0.5$ in $2/1$ resonant capture and for $\rho=0.5$ where $K=0.5$ in $3/1$ resonant capture). The increase of mutual inclination occurs, when the system reaches a {\em vco}.} 
\label{FigMig3D21}
\end{figure}

\begin{figure}
$$
\begin{array}{cc}
\includegraphics[width=6cm]{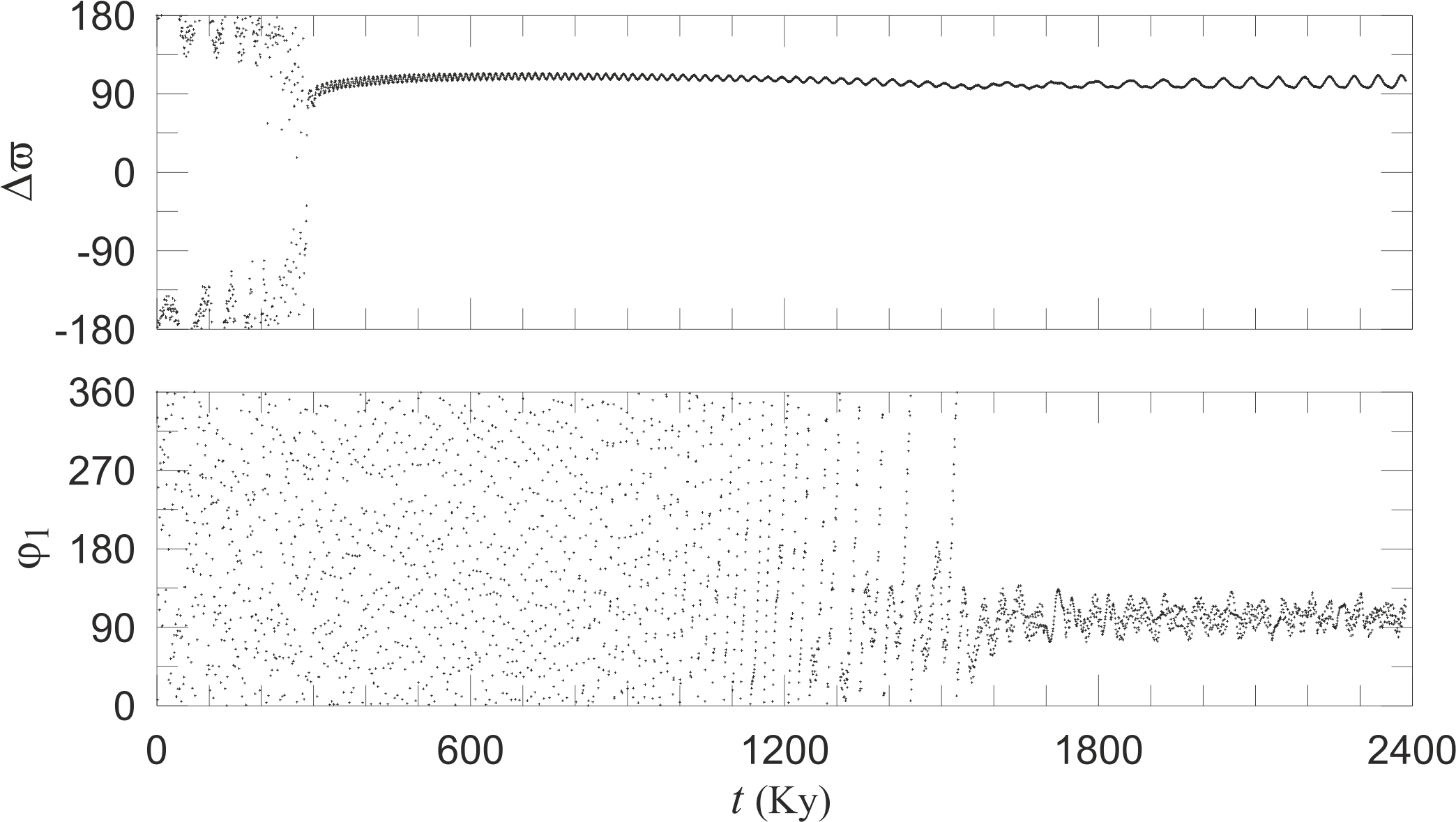}  & \includegraphics[width=6cm]{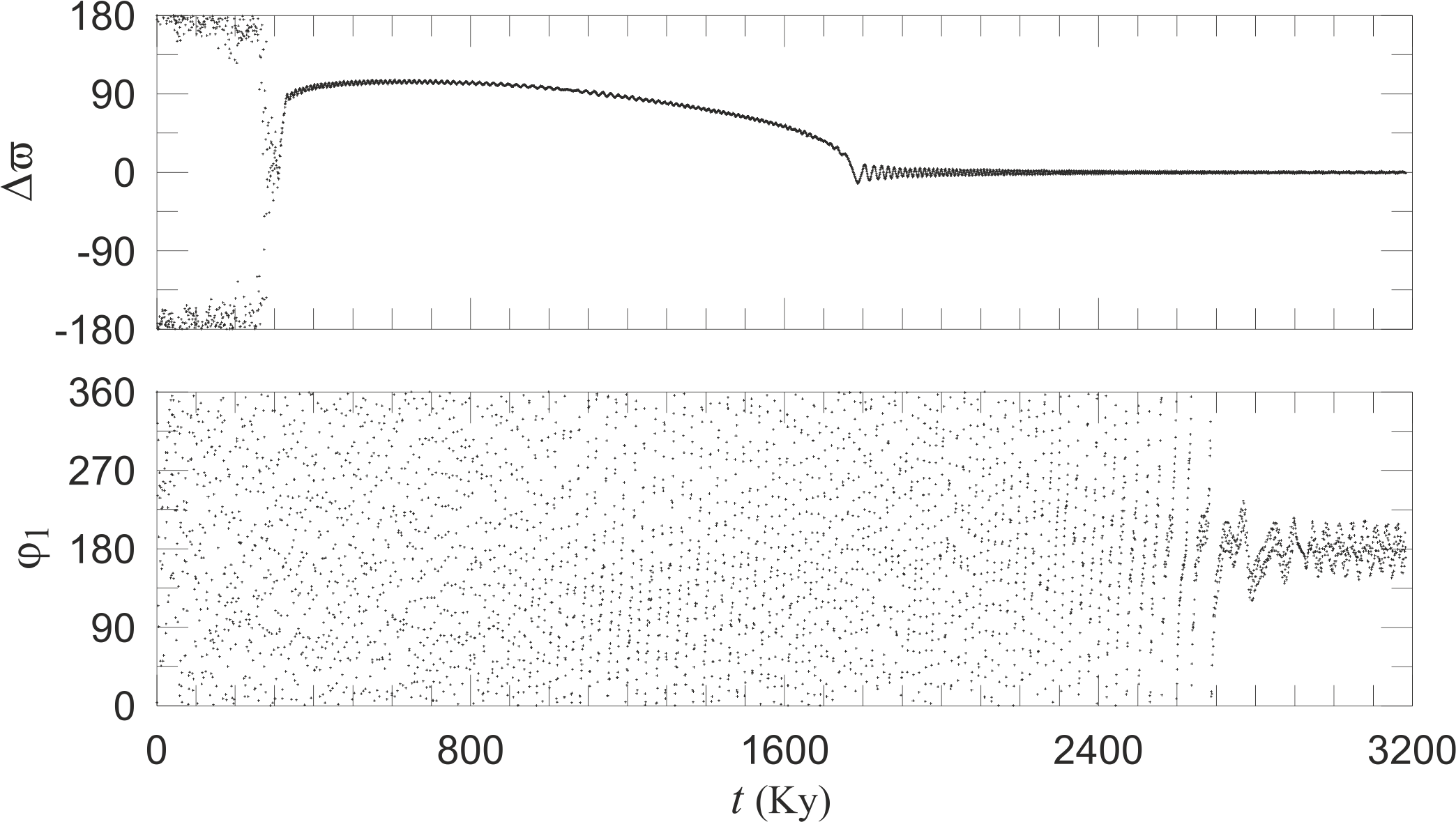} \\
\textnormal{(a) 2/1,}\; \rho=0.4  &\textnormal{(b) 2/1,}\; \rho=0.5 \\
\includegraphics[width=6cm]{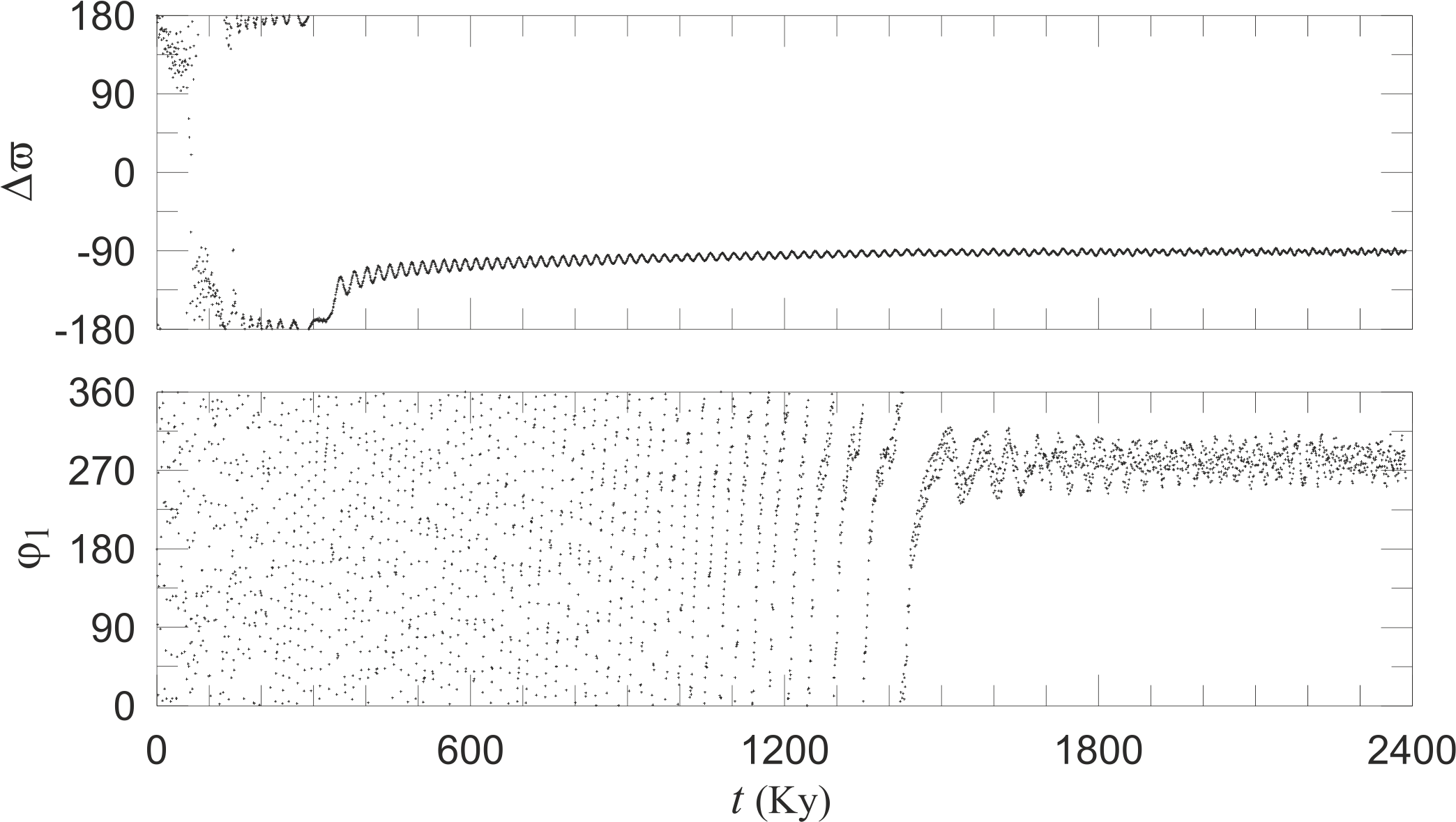}  & \includegraphics[width=6cm]{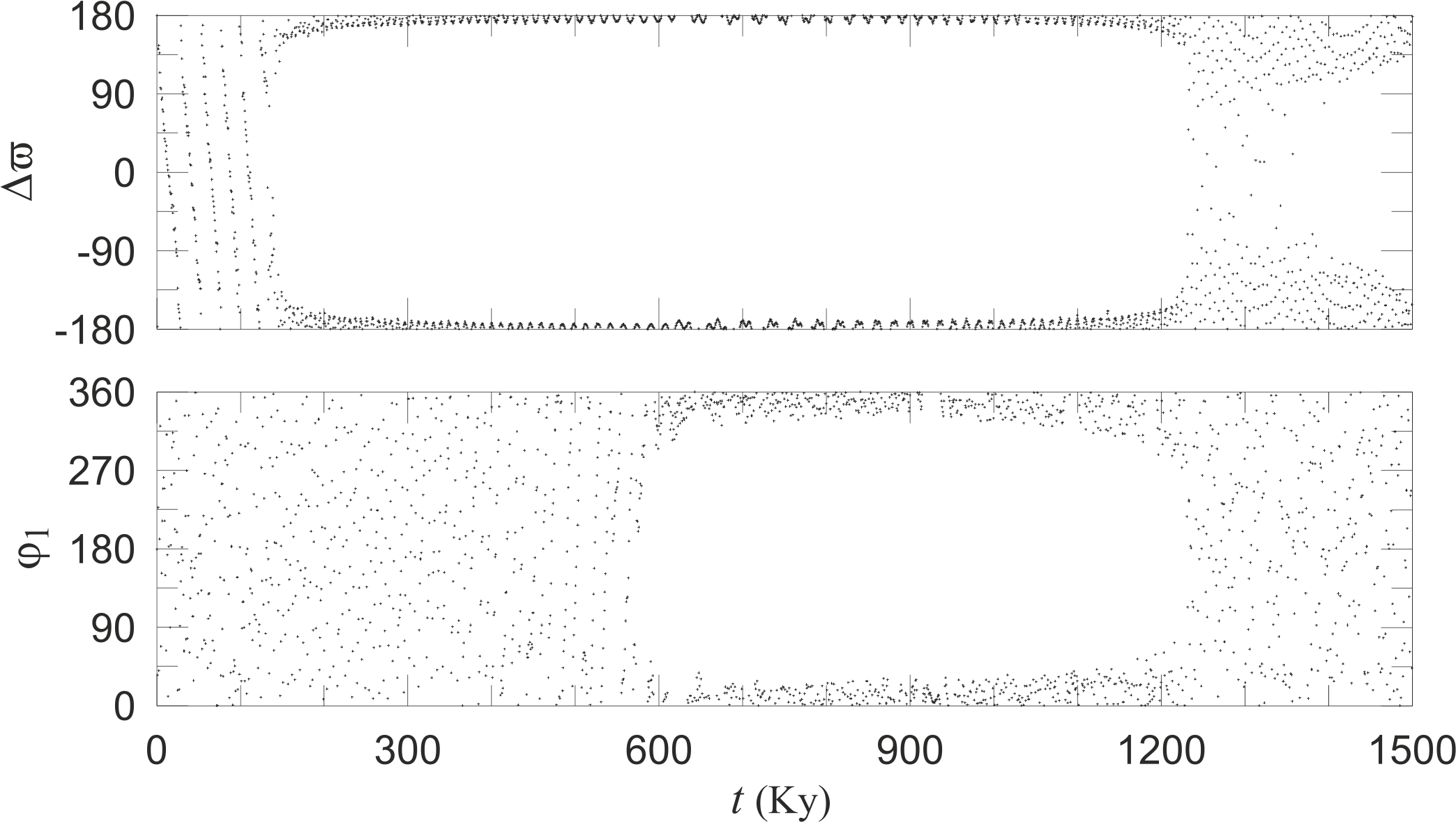} \\
\textnormal{(c) 3/1,}\; \rho=1.0  &\textnormal{(d) 3/1,}\; \rho=10.0  
\end{array}$$
\caption{The evolution of the resonant angles $\Delta\varpi$ and $\varphi_1$, for the indicated resonance and mass ratio $\rho=m_2/m_1$ and for $\nu=10^{-6}\,y^{-1}$, $K=1$. After the capture in the resonance $\Delta\varpi$ librates and after the inclination resonance $\varphi_1$ also librates. Librations take place around $0^\circ$ or $180^\circ$ (symmetric configuration) or around a different value (asymmetric configuration).} 
\label{FigRAsas}
\end{figure}

The different migration paths, associated with different mass ratios as was discussed above, are presented in Fig. \ref{FigMig3D21}a. In all cases the system enters the inclination resonance, when the {\em vco} is reached. The different loci of the ``inclination resonance'' for $\rho<\rho^*$ and $\rho>\rho^*$ are clearly  distinguished.  In Fig. \ref{FigRAsas} (panels a, b), we present the evolution of resonant angles $\Delta\varpi$ and $\varphi_1$ for $\rho=0.4<\rho^*$ and $\rho=0.5>\rho^*$. After the capture in the resonance $\Delta\varpi$ librates and when the system enters the inclination resonance $\varphi_1$ also librates. In the first case (panel a), we finally obtain an asymmetric libration for both resonant angles. In the second case (panel b), after a passage from asymmetric librations for $\Delta\varpi$ (along the family $A_a^{2/1}$), both angles, $\Delta\varpi$ and $\varphi_1$ librate around $0^\circ$ and $180^\circ$, respectively. So the bifurcation at $\rho=\rho^*$ explains the result of \citet{leetho09} that at $\rho\approx 0.4$ the inclination resonance changes from asymmetric to symmetric.  

The above types of evolution have been always verified by our simulations for $\nu\lesssim 10^{-6}\,y^{-1}$ and for small eccentricity damping ($K\approx 1$) that permits the sufficient increase of the eccentricities. For larger values of $\nu$, particularly for $\nu=10^{-5}\,y^{-1}$, we obtained capture in the 2/1 resonant and in most cases temporal inclination resonance followed by large oscillation of the eccentricities, which in a relatively short time interval destabilizes the system and the evolution becomes strongly irregular. We observed that it is possible even for $\rho<\rho^*$ the evolution to overcome the asymmetric {\em vco} and then either the inclination resonance appears when the system reaches the next {\em vco} located in the symmetric family or the system jumps in an other family of periodic orbits. Such complicated phenomena, which are possible for reasonable values of the migration rate, have been noticed and discussed in \citet{leetho09}.                       

\subsection{$3/1$ resonance capture}
In the $3/1$ resonance, the system migrates along the family $S_4^{3/1}$. When this family becomes horizontally unstable, the system follows the asymmetric family $A_4^{3/1}$, which starts both as horizontally and vertically stable (see Fig. \ref{FigEfam21}b). As we have mentioned, along family $A_4^{3/1}$ a {\em vco} exists, for any mass ratio value. There the family becomes vertically unstable and the inclination resonance occurs. Fig. \ref{FigMig3D21}b shows the evolution in the space $e_1-e_2-\Delta i$ for different values of $\rho$. An increase of the mutual inclination occurring when the {\em vco} is reached is clearly seen in all cases. Also, in Fig. \ref{FigRAsas} (panels c,d), we present the evolution of resonant angles $\Delta\varpi$ and $\varphi_1$=$\lambda_1-3\lambda_2+2\Omega_1$. For $m_1=m_2$ (panel c) $\Delta\varpi$ librates around $180^\circ$ after the 3/1 resonance capture in family $S_4^{3/1}$ and then its librations become asymmetric (bifurcation to the family $A_4^{3/1}$). Angle $\varphi_1$ starts to librate around $290^\circ$, when the system reaches the {\em vco}. The inclination resonance becomes symmetric for $\rho=10$ \footnote{Particularly we used $m_1=0.0005$ and $m_2=0.005$. We remind that for $\rho\gtrsim6.5$, the {\em vco} belongs to the stable part of the family $S_4^{3/1}$}, as it is shown in panel (d), where $\varphi_1$ librates around $0^\circ$ after 600 Ky, where the inclination resonance takes place. However, after about 1.2 My the evolution becomes quite irregular and the angles show rotations.    

Differential migration with positive eccentricity damping generally seems to reach an asymptotic limit at some point in the $(e_1,e_2)$ plane (\citealt{leepeal02,kley03}). This point depends on the values of the migration rate, $\nu$, and eccentricity damping rate, $K$. In the case of $\rho=0.5$ shown in Fig. \ref{FigMig3D21}b, we have $K=0.5$ (in contrast to the remaining cases, where $K=1$). If for the same initial conditions we also used $K=1$, then the system would stall at $(e_1,e_2)\approx(0.15,0.4)$ and therefore, the {\em vco} of this particular family would not be reached and inclination excitation would not occur. The same situation holds also in the $2/1$ resonance (see Fig. \ref{FigMig3D21}a) for $\rho=0.3$.   

Our numerical simulations showed that the capture in the 3/1 resonance requires quite slower migration rates compared to those for the 2/1 case and depends significantly on the planetary mass ratio, $\rho$. Particularly, for $K=0.5$ and $\nu=10^{-6}\,y^{-1}$ or $\nu=5\times 10^{-7}\,y^{-1}$ capture in the 3/1 resonance is observed only for $\rho\lesssim 0.8$ or $\rho\lesssim 0.5$, respectively. When the system is captured in the 3/1 resonance, then the paths defined by the particular families of periodic orbits are followed by the evolution and inclination resonance always occurs, if the eccentricity damping is sufficiently small ($K\lesssim 1$) for the system to reach the {\em vco}.

\section{Conclusions and discussion}
Previous studies have shown that planet migration induced by tidal interactions between Jovian-sized planets and the gaseous protoplanetary disc, takes place along specific paths in phase space, formed by stable families of periodic orbits. In this paper, we considered the spatial case of the three-body problem, in order to describe possible effects of migration on the mutual inclination of two-planet systems.
  
We computed the vertical stability index along families of planar periodic orbits and determined the vertical critical orbits ({\em vco}). Orbits with initial conditions in the neighbourhood of a horizontally and vertically stable periodic orbit, evolve regularly, showing very small oscillations in the inclinations. However, in the neighbourhood of vertically unstable periodic orbits, the initially small mutual planetary inclination may increase to high values, depending on the mass ratio, as well as the mean motion ratio of the two planets.    

We showed that the ``inclination resonance'', which was observed in the numerical simulations of \citet{thommes03}, \citet{leetho09} and \citet{litsi09b}, should be associated with the existence of {\em vco}'s along the corresponding planar family of resonant periodic orbits. In particular, when an initially almost planar system migrates along a horizontally and vertically stable family of periodic orbits, the initial small inclinations show oscillations of very small amplitude. When, a {\em vco} is reached the inclinations start to increase rapidly. The position of a {\em vco} along a family of periodic orbits depends on the planetary mass ratio, $\rho$.

\begin{figure}
$$
\begin{array}{cc}
\includegraphics[width=6cm]{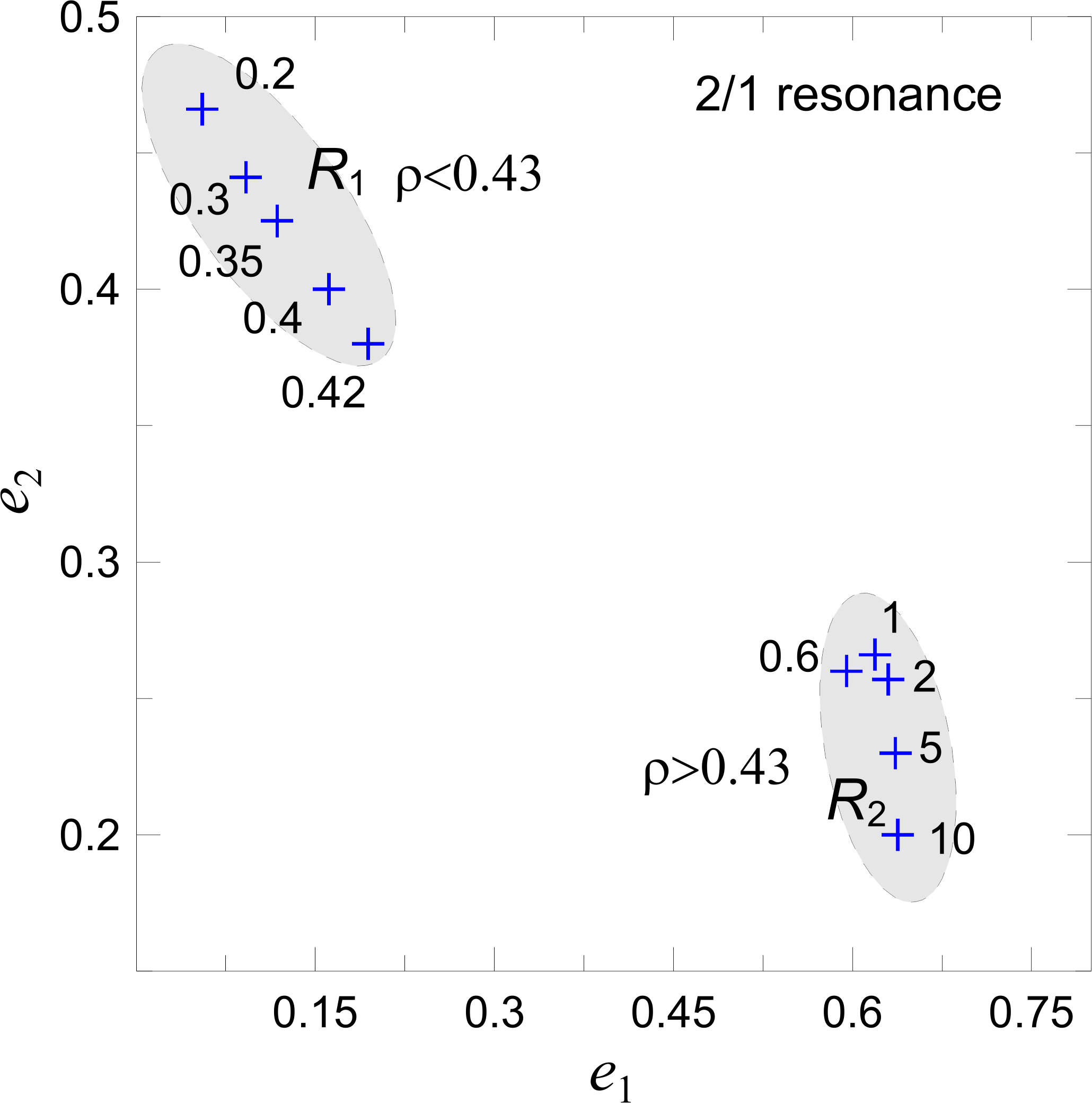}  & \includegraphics[width=6cm]{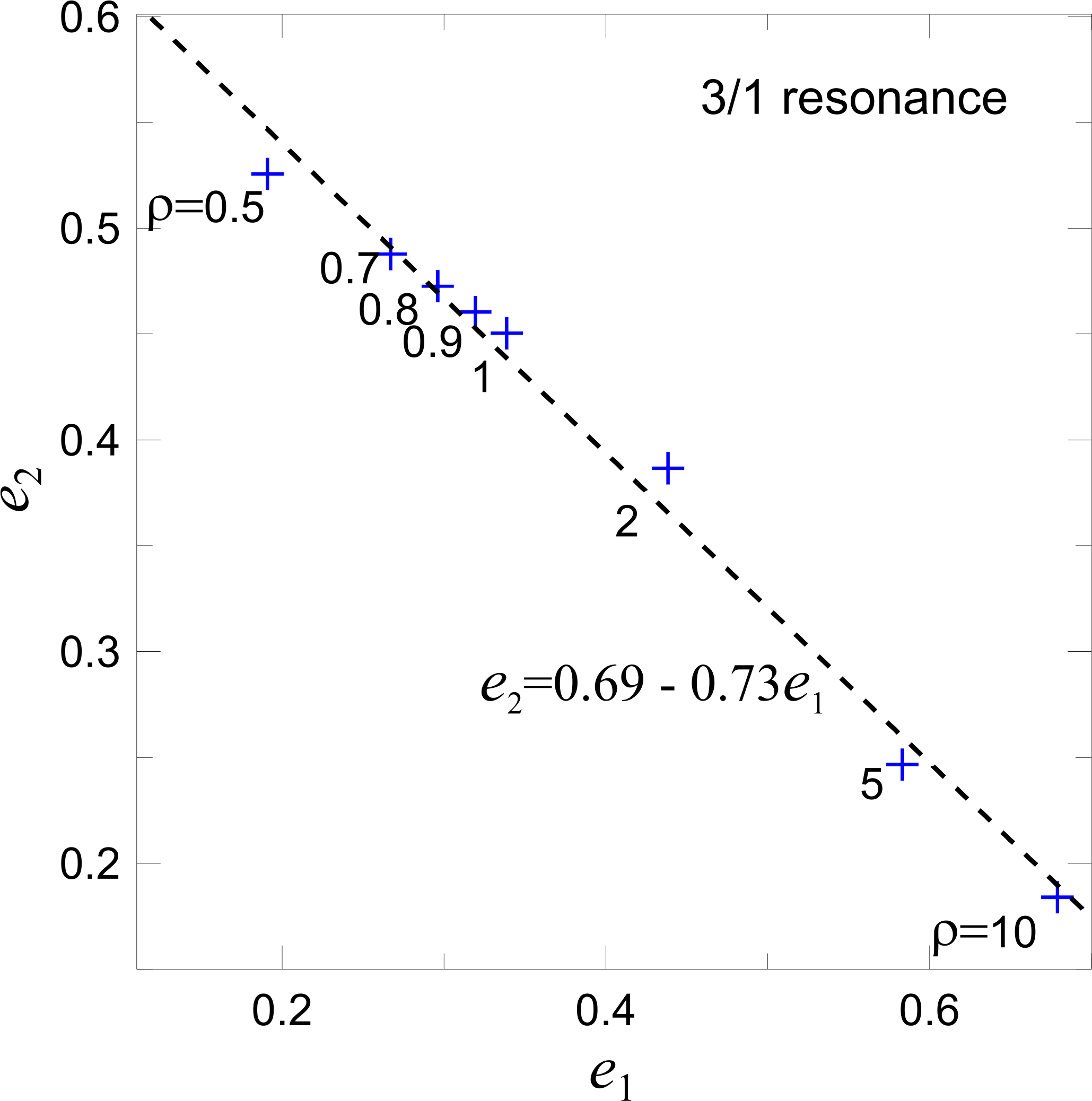} \\
\textnormal{(a)} &\textnormal{(b)}  
\end{array}$$
\caption{The position of {\em vco} related to inclination resonance after differential migration. The corresponding planetary mass ratio, $\rho$, is indicated for each {\em vco} {\bf a} $2/1$ resonance: two distinct regions of {\em vco} are obtained  {\bf b} $3/1$ resonance: the {\em vco} are located on a straight line in the plane of eccentricities (the least square fitting is presented).} 
\label{FigVcoDistr}
\end{figure}

We particularly studied the $2/1$ and $3/1$ resonant captures. The families of periodic orbits for these resonances (and various values of $\rho$) have {\em vco}'s, which can give inclination excitation, when they are reached by differential migration. The distribution of these {\em vco}'s in the eccentricities plane is different for these two resonances and is presented in Fig. \ref{FigVcoDistr}. For the $2/1$ resonance, we obtain two distinct regions of {\em vco}'s, $R_1$ and $R_2$. In region $R_1$, the eccentricity of the outer planet is larger than the inner one. This region can be reached by a migrating system with mass ratio $\rho<0.43$. For larger mass ratios, the system reaches the region $R_2$, where the orbit of the inner planet is quite more eccentric than the orbit of the outer one. 

In the $3/1$ resonance, the {\em vco}'s seem to be located approximately on a straight line in the plane of eccentricities. As $\rho$ increases, $e_1$ increases and $e_2$ decreases. This relationship can be expressed by a linear fit of the form $e_2=0.69-0.73e_1$. For $\rho\lesssim 6.5$, the {\em vco}'s that drive migration belong to the asymmetric families of periodic orbits, while for larger mass ratio values, they belong to the symmetric families, $S_4^{3/1}$. 

We note that a similar relation between the onset of inclination excitation and the values of the planetary eccentricities was pointed out by \citet{litsi09b}, who suggested that, for inclination excitation to occur, at least one of the two planets has to have an eccentricity larger than $\gtrsim 0.4$. Here, we further quantify this relationship. Moreover, we show that it is a direct consequence of the distribution of {\em vco}'s in the eccentricities plane, for different mass ratios. Our results offer a possible diagnostic tool for $2/1$- and $3/1$-resonant systems: if the mass ratio and the eccentricities are known, then, we can tell if the system had passed through a {\em vco} and thus, would have a non-zero mutual inclination.

Differential migration generally seems to stall, reaching an asymptotic limit in $(e_1,e_2)$; beyond this point in time no further increase of the eccentricities occurs. This limit and the time that the systems takes to reach it depend on the migration parameters. In the $2/1$ and $3/1$ resonances studied in this paper, generally the {\em vco}'s appear for relatively high eccentricity values of one of the two planets. Thus, a necessary condition for the inclination resonance to occur is that the differential migration can be sustained for long-enough times and is sufficiently `strong', as to bring the system at the {\em vco} position. As our numerical results suggest, this will be true for small values of the eccentricity damping rate, $K$. When fast eccentricity variations along migration cause `jumps' to other regions in phase space, whence capture to different families occurs, the presence of vertical instability in these families should excite the inclinations.

In our model we considered a simplified model to describe the interaction of the planets with the gaseous protoplanetary disc, which is better suited for a planar model, rather than a $3D$ one. When the system becomes inclined, the dissipative forces assumed here (\ref{Fdissipative}) do not adequately describe the averaged planet-disc interactions. Thus, although we are confident about the occurrence of vertical instability on the {\em vco}'s, we cannot safely conclude what are the maximum values of mutual inclination that a particular system could reach.

{\bf Acknowledgments.} This research has been co-financed by the European Union (European Social Fund - ESF) and Greek national funds through the Operational Program ``Education and Lifelong Learning'' of the National Strategic Reference Framework (NSRF) - Research Funding Program: Thales. Investing in knowledge society through the European Social Fund. The work of K.T.~ was supported by AUTh Research Committee's ``Action C: Support of Research Activities in Basic Research'' (Contract Nr.~ 89406). 

\bibliographystyle{plainnat}
\bibliography{bib}

\begin{thebibliography}{35}
\providecommand{\natexlab}[1]{#1}
\providecommand{\url}[1]{\texttt{#1}}
\expandafter\ifx\csname urlstyle\endcsname\relax
  \providecommand{\doi}[1]{doi: #1}\else
  \providecommand{\doi}{doi: \begingroup \urlstyle{rm}\Url}\fi

\bibitem[Antoniadou and Voyatzis(2013)]{av12}
K.~I. Antoniadou and G.~Voyatzis.
\newblock 2/1 resonant periodic orbits in three dimensional planetary systems.
\newblock \emph{Celestial Mechanics and Dynamical Astronomy}, 115:\penalty0
  161--184, 2013.

\bibitem[Beaug{\'e} and Ferraz-Mello(1993)]{bf93}
C.~Beaug{\'e} and S.~Ferraz-Mello.
\newblock Resonance trapping in the primordial solar nebula - the case of a
  stokes drag dissipation.
\newblock \emph{Icarus}, 103:\penalty0 301--318, 1993.

\bibitem[Beaug{\'e} et~al.(2006)Beaug{\'e}, Michtchenko, and
  Ferraz-Mello]{bmfm06}
C.~Beaug{\'e}, T.~A. Michtchenko, and S.~Ferraz-Mello.
\newblock Planetary migration and extrasolar planets in the 2/1 mean-motion
  resonance.
\newblock \emph{Monthly Notices of the Royal Astronomical Society},
  365:\penalty0 1160--1170, 2006.

\bibitem[{Correa-Otto} et~al.(2013){Correa-Otto}, {Michtchenko}, and
  {Beaug{\'e}}]{otto13}
J.~A. {Correa-Otto}, T.~A. {Michtchenko}, and C.~{Beaug{\'e}}.
\newblock {A new scenario for the origin of the 3/2 resonant system HD 45364}.
\newblock \emph{Astronomy and Astrophysics}, 560:\penalty0 A65, 2013.

\bibitem[Ferraz-Mello et~al.(2003)Ferraz-Mello, Beaug{\'e}, and
  Michtchenko]{mebeaumich03}
S.~Ferraz-Mello, C.~Beaug{\'e}, and T.~A. Michtchenko.
\newblock Evolution of migrating planet pairs in resonance.
\newblock \emph{Celestial Mechanics and Dynamical Astronomy}, 87:\penalty0
  99--112, 2003.

\bibitem[Ferraz-Mello et~al.(2005)Ferraz-Mello, Michtchenko, and
  Beaug{\'e}]{fmb05}
S.~Ferraz-Mello, T.~A. Michtchenko, and C.~Beaug{\'e}.
\newblock The orbits of the extrasolar planets hd 82943c and b.
\newblock \emph{The Astrophysical Journal}, 621:\penalty0 473--481, 2005.

\bibitem[Hadjidemetriou(2006)]{hadj06}
J.~D. Hadjidemetriou.
\newblock Symmetric and asymmetric librations in extrasolar planetary systems:
  a global view.
\newblock \emph{Celestial Mechanics and Dynamical Astronomy}, 95:\penalty0
  225--244, 2006.

\bibitem[Hadjidemetriou and Voyatzis(2010)]{hadjvoy10}
J.~D. Hadjidemetriou and G.~Voyatzis.
\newblock On the dynamics of extrasolar planetary systems under dissipation:
  Migration of planets.
\newblock \emph{Celestial Mechanics and Dynamical Astronomy}, 107:\penalty0
  3--19, 2010.

\bibitem[Hadjidemetriou and Voyatzis(2011)]{hv11}
J.~D. Hadjidemetriou and G.~Voyatzis.
\newblock Different types of attractors in the three body problem perturbed by
  dissipative terms.
\newblock \emph{International Journal of Bifurcation and Chaos}, 21:\penalty0
  2195--2209, 2011.

\bibitem[Haghighipour(1999)]{haghi99}
N.~Haghighipour.
\newblock Dynamical friction and resonance trapping in planetary systems.
\newblock \emph{Monthly Notices of the Royal Astronomical Society},
  304:\penalty0 185--194, 1999.

\bibitem[H\'enon(1973)]{hen}
M.~H\'enon.
\newblock Vertical stability of periodic orbits in the restricted problem. i.
  equal masses.
\newblock \emph{Astronomy and Astrophysics}, 28:\penalty0 415, 1973.

\bibitem[{Henrard}(1982)]{henr82}
J.~{Henrard}.
\newblock {Capture into resonance - an extension of the use of adiabatic
  invariants}.
\newblock \emph{Celestial Mechanics}, 27:\penalty0 3--22, 1982.

\bibitem[{Henrard} and {Lemaitre}(1983)]{henrl83}
J.~{Henrard} and A.~{Lemaitre}.
\newblock {A second fundamental model for resonance}.
\newblock \emph{Celestial Mechanics}, 30:\penalty0 197--218, 1983.

\bibitem[Ichtiaroglou and Michalodimitrakis(1980)]{ichmich80}
S.~Ichtiaroglou and M.~Michalodimitrakis.
\newblock Three-body problem - the existence of families of three-dimensional
  periodic orbits which bifurcate from planar periodic orbits.
\newblock \emph{Astronomy and Astrophysics}, 81:\penalty0 30--32, 1980.

\bibitem[Ichtiaroglou et~al.(1978)Ichtiaroglou, Katopodis, and
  Michalodimitrakis]{ikm78}
S.~Ichtiaroglou, K.~Katopodis, and M.~Michalodimitrakis.
\newblock On the continuation of periodic orbits in the three-body problem.
\newblock \emph{Astronomy and Astrophysics}, 70:\penalty0 531, 1978/11/1 1978.

\bibitem[Kley(2003)]{kley03}
W.~Kley.
\newblock Dynamical evolution of planets in disks.
\newblock \emph{Celestial Mechanics and Dynamical Astronomy}, 87:\penalty0
  85--97, 2003.

\bibitem[Lee(2004)]{lee04}
M.~H. Lee.
\newblock Diversity and origin of 2:1 orbital resonances in extrasolar
  planetary systems.
\newblock \emph{The Astrophysical Journal}, 611:\penalty0 517--527, 2004.

\bibitem[Lee and Peale(2002)]{leepeal02}
M.~H. Lee and S.~J. Peale.
\newblock Dynamics and origin of the 2:1 orbital resonances of the gj 876
  planets.
\newblock \emph{The Astrophysical Journal}, 567:\penalty0 596--609, 2002.

\bibitem[Lee and Thommes(2009)]{leetho09}
M.~H. Lee and E.~W. Thommes.
\newblock Planetary migration and eccentricity and inclination resonances in
  extrasolar planetary systems.
\newblock \emph{The Astrophysical Journal}, 702:\penalty0 1662--1672, 2009.

\bibitem[Libert and Tsiganis(2009)]{litsi09b}
A.-S. Libert and K.~Tsiganis.
\newblock Trapping in high-order orbital resonances and inclination excitation
  in extrasolar systems.
\newblock \emph{Monthly Notices of the Royal Astronomical Society},
  400:\penalty0 1373--1382, 2009.

\bibitem[Michalodimitrakis(1979)]{mich79}
M.~Michalodimitrakis.
\newblock On the continuation of periodic orbits from the planar to the
  three-dimensional general three-body problem.
\newblock \emph{Celestial Mechanics}, 19:\penalty0 263--277, 1979/4/1 1979.

\bibitem[{Michtchenko} and {Rodr{\'{\i}}guez}(2011)]{mr11}
T.~A. {Michtchenko} and A.~{Rodr{\'{\i}}guez}.
\newblock {Modelling the secular evolution of migrating planet pairs}.
\newblock \emph{Monthly Notices of the Royal Astronomical Society},
  415:\penalty0 2275--2292, 2011.

\bibitem[Michtchenko et~al.(2006)Michtchenko, Beaug{\'e}, and
  Ferraz-Mello]{mbf06}
T.~A. Michtchenko, C.~Beaug{\'e}, and S.~Ferraz-Mello.
\newblock Stationary orbits in resonant extrasolar planetary systems.
\newblock \emph{Celestial Mechanics and Dynamical Astronomy}, 94:\penalty0
  411--432, 2006.

\bibitem[{Morbidelli} and {Crida}(2007)]{morc07}
A.~{Morbidelli} and A.~{Crida}.
\newblock {The dynamics of Jupiter and Saturn in the gaseous protoplanetary
  disk}.
\newblock \emph{Icarus}, 191:\penalty0 158--171, 2007.

\bibitem[{Morbidelli} et~al.(2009){Morbidelli}, {Brasser}, {Tsiganis}, {Gomes},
  and {Levison}]{morbtgl09}
A.~{Morbidelli}, R.~{Brasser}, K.~{Tsiganis}, R.~{Gomes}, and H.~F. {Levison}.
\newblock {Constructing the secular architecture of the solar system. I. The
  giant planets}.
\newblock \emph{Astronomy and Astrophysics}, 507:\penalty0 1041--1052, 2009.

\bibitem[Nelson and Papaloizou(2002)]{np02}
R.~P. Nelson and J.~C.~B. Papaloizou.
\newblock Possible commensurabilities among pairs of extrasolar planets.
\newblock \emph{Monthly Notices of the Royal Astronomical Society},
  333:\penalty0 L26--L30, 2002.

\bibitem[Papaloizou(2003)]{pap03}
J.~C.~B. Papaloizou.
\newblock Disc-planet interactions: Migration and resonances in extrasolar
  planetary systems.
\newblock \emph{Celestial Mechanics and Dynamical Astronomy}, 87:\penalty0
  53--83, 2003.

\bibitem[{Peale}(1986)]{peale86}
S.~J. {Peale}.
\newblock \emph{{Orbital resonances, unusual configurations and exotic rotation
  states among planetary satellites}}, pages 159--223.
\newblock University of Arizona Press, 1986.

\bibitem[Skokos(2001)]{skokos01}
C.~Skokos.
\newblock On the stability of periodic orbits of high dimensional autonomous
  hamiltonian systems.
\newblock \emph{Physica D Nonlinear Phenomena}, 159:\penalty0 155--179, 2001.

\bibitem[Thommes and Lissauer(2003)]{thommes03}
E.~W. Thommes and J.~J. Lissauer.
\newblock Resonant inclination excitation of migrating giant planets.
\newblock \emph{The Astrophysical Journal}, 597:\penalty0 566--580, November
  2003.

\bibitem[Voyatzis(2008)]{voyatzis08}
G.~Voyatzis.
\newblock Chaos, order, and periodic orbits in 3:1 resonant planetary dynamics.
\newblock \emph{The Astrophysical Journal}, 675:\penalty0 802--816, 2008.

\bibitem[Voyatzis and Hadjidemetriou(2005)]{voyhadj05}
G.~Voyatzis and J.~D. Hadjidemetriou.
\newblock Symmetric and asymmetric librations in planetary and satellite
  systems at the 2/1 resonance.
\newblock \emph{Celestial Mechanics and Dynamical Astronomy}, 93:\penalty0
  263--294, 2005.

\bibitem[Voyatzis and Hadjidemetriou(2006)]{vh06}
G.~Voyatzis and J.~D. Hadjidemetriou.
\newblock Symmetric and asymmetric 3:1 resonant periodic orbits with an
  application to the 55cnc extra-solar system.
\newblock \emph{Celestial Mechanics and Dynamical Astronomy}, 95:\penalty0
  259--271, 2006.

\bibitem[Voyatzis et~al.(2009)Voyatzis, Kotoulas, and Hadjidemetriou]{vkh09}
G.~Voyatzis, T.~Kotoulas, and J.~D. Hadjidemetriou.
\newblock On the 2/1 resonant planetary dynamics - periodic orbits and
  dynamical stability.
\newblock \emph{Monthly Notices of the Royal Astronomical Society},
  395:\penalty0 2147--2156, 2009.

\bibitem[{Ward}(1997)]{war97}
W.~R. {Ward}.
\newblock {Protoplanet Migration by Nebula Tides}.
\newblock \emph{Icarus}, 126:\penalty0 261--281, 1997.

\end{thebibliography}
\end{document}